\documentclass[11pt]{article}

\usepackage[margin=1in]{geometry}
\usepackage{amsmath,amssymb}
\usepackage{graphicx}
\usepackage{booktabs}
\usepackage{subcaption}
\usepackage[super,sort&compress,comma]{natbib}
\usepackage{microtype}
\usepackage{xcolor}
\usepackage{hyperref}
\usepackage{setspace}
\newcommand{\tv}{\mathrm{TV}}
\newcommand{\Sim}{\mathrm{cos}}

\title{\textbf{Large language models simulate intersectional synthetic identities with a budget of one to two dimensions}}
\author{
Virgile Rennard\thanks{Massachusetts Institute of Technology (MIT).}
\and
Christos Xypolopoulos\thanks{\'Ecole Polytechnique.}
}
\date{}

\begin{document}
\maketitle

\begin{abstract}

Large language models are increasingly used as synthetic survey respondents, promising cheap access to rare intersectional populations. We test standard demographic-persona methods against every real intersectional subgroup (n $\geq$ 20) across 15 waves of Pew's American Trends Panel—21 million simulated response distributions from eight models. In real respondents, subgroup opinion is approximately the additive sum of its single-identity components, yet grows 2.5× more distinctive as identities intersect. Simulated respondents show no such composition: a single feature explains a two-feature persona's responses better than the additive combination in 75–82\% of subgroups, and a third feature adds almost nothing. This collapse survives every prompting strategy we test. Additionally, the feature models retain is chosen nearly blindly—except that they systematically discard race and religion, the strongest real drivers of opinion. Synthetic samples offer intersectional personas but represent one identity at a time.

\end{abstract}

\section*{Introduction}

A single wave of a major US probability panel contains roughly thirty
Black Republicans. ``Synthetic identities'', prompting a large language model
(LLM) to answer surveys as a member of a demographic
group\citep{argyle2023out,sun2024random,park2024generative}, promise a
thousand of them for the cost of an API call, and the promise is most
valuable exactly where real data is thinnest: small intersectional
populations carrying several identity-relevant attributes at
once\citep{crenshaw1989demarginalizing}. Whether to trust it is a live
methodological debate\citep{dillion2023can,grossmann2023ai,messeri2024artificial,
bisbee2024synthetic} that existing audits do not answer: evaluations
score one attribute at a time\citep{santurkar2023whose,hwang2023aligning},
or score joint personas only against the joint
target\citep{argyle2023out,bisbee2024synthetic}, while the fairness
literature examines open-ended text rather than measurable opinion
distributions\citep{cheng2023compost,liu2024evaluating,wang2025large}.
To our knowledge, no prior work tests the \emph{compositional} question
at distribution level against real intersectional ground truth: whether a
model told that a persona is Black \emph{and} Republican integrates both
facts, or silently discards one. Composition is not an edge case but the
operative requirement for simulating any real population, because real
people are never one attribute.

This paper measures composition directly. We prompt eight LLMs---7B open-weights models through two current-generation frontier systems---to produce full response distributions for 1{,}000 hypothetical respondents matching profiles of one, two, or three features from seven demographic dimensions, for every question in 15 waves of the Pew American Trends Panel (ATP): 15.7 million simulated distributions in the primary pipeline, 21.1 million across all conditions. The ground truth is itself intersectional: from respondent-level microdata we computed the real response distribution of every demographic cell with at least 20 respondents, up to three-way intersections, so each simulation is scored against the measured behaviour of exactly the population it claims to represent. Throughout, ``intersectional'' and ``additivity'' are operationalised deliberately conservatively, at the level of ATP closed-form \emph{marginal opinion distributions}: our additivity benchmark asks only whether two subgroups' measured response distributions combine, and neither captures nor contradicts the structural and experiential account of intersectionality\citep{crenshaw1989demarginalizing}. 

We validate the additivity not as an assumption but a property: real ATP subgroups turn out to be near-additive in both the direction and the magnitude of their opinion tilts (Fig.~\ref{fig:humancontest}), so the benchmark the models face is approximately what real populations do---and a model that fails even this distributional, statistically conservative test cannot be capturing the richer phenomenon.

Three methodological safeguards, which we suggest as standards for any silicon-sampling audit, shape every result. First, an explicit \emph{human sampling-noise floor}: small real cells have noisy ground truth, and we show this noise manufactures spurious findings if uncorrected. Second, a \emph{null calibration} for similarity metrics: bias vectors from one model share a global error component, so a random unrelated dimension already ``explains'' a profile's bias at median cosine 0.84---absolute similarities are uninterpretable without this floor. Third, full \emph{split-sample confirmation}: every headline number replicates on disjoint halves of the survey set.

The findings form a single pattern. Models collapse multi-feature personas onto one to two dimensions; the collapse direction is near-blind to which feature actually matters; the strongest real drivers (race, religion) are the most reliably discarded; added identity information buys no accuracy once noise is controlled; and failure is uniform across profile typicality, leaving no safe subset of intersections. The numbers are strikingly stable across eight architectures from five organizations. Real populations move the other way, growing measurably more distinctive as identities intersect. The gap between those facts is the empirical case against synthetic intersectional participants.

\begin{figure}[h!]
\centering
\includegraphics[width=.95\textwidth]{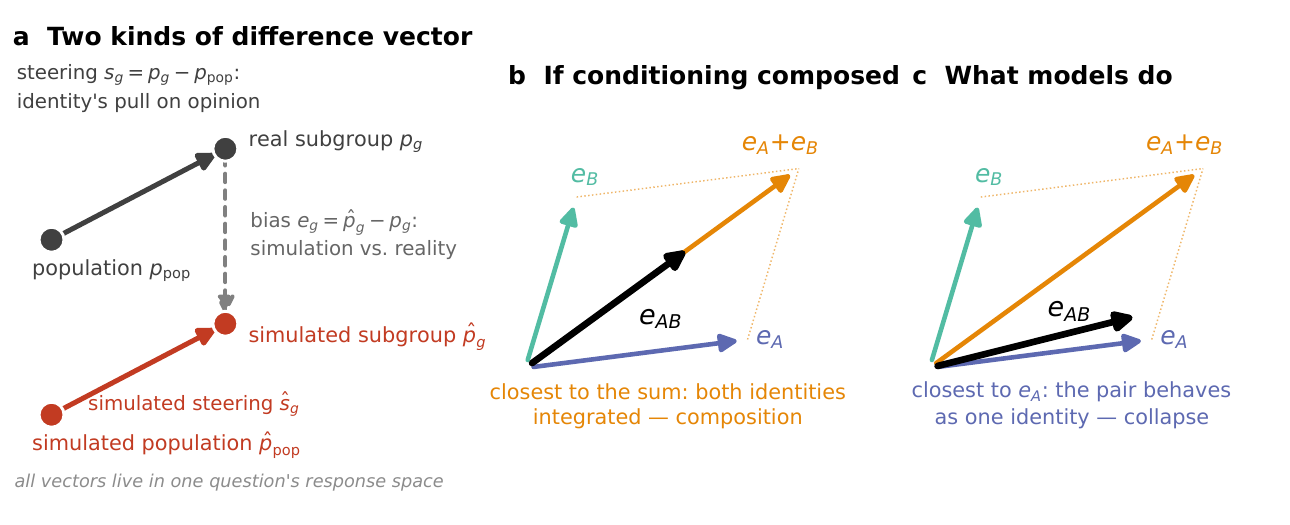}
\caption{\textbf{The measurement machinery.}
\textbf{a}, For each demographic cell $g$ and question, two difference
vectors are formed inside that question's response space:
\emph{steering}, $s_g = p_g - p_{\mathrm{pop}}$, measures how holding an
identity moves a distribution away from the population (real and
simulated versions); \emph{bias}, $e_g = \hat p_g - p_g$, measures how
far the simulation sits from reality for the same cell. \textbf{b--c},
The composition contest for a pair profile $AB$: the model's own
single-feature biases supply two candidate explanations of the realized
pair bias $e_{AB}$---their sum (additive; both identities integrated)
and each single alone---scored by cosine similarity. If conditioning
composed as it does in the \textbf{ground truth}, $e_{AB}$ would track the sum (\textbf{b}); across eight
models it instead tracks one single feature (\textbf{c}): collapse.
Raw win rates from this contest are never interpreted directly; they
are calibrated against synthetic cells in which the truth is known
(Methods).}
\label{fig:schematic}
\end{figure}

\begin{figure}[h!]
\centering
\begin{subfigure}{0.58\textwidth}
  \includegraphics[width=\linewidth]{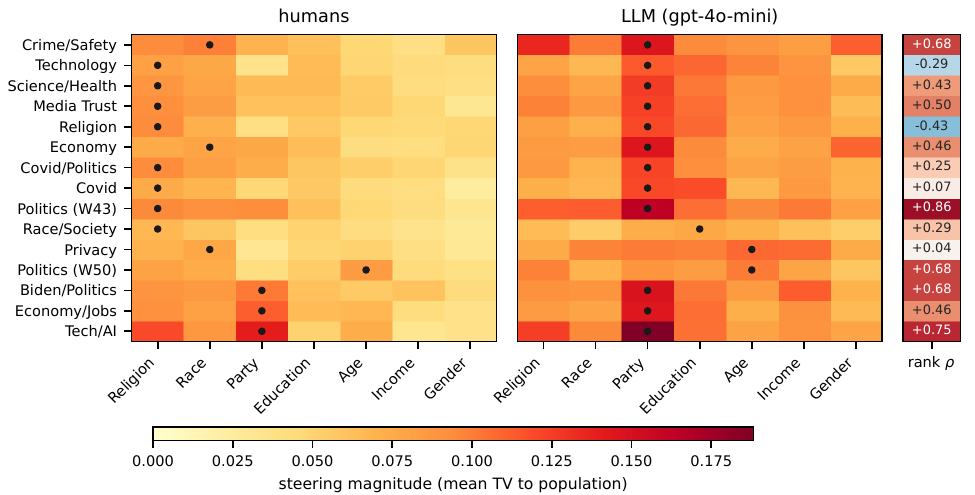}
  \caption{}
\end{subfigure}\hfill
\begin{subfigure}{0.40\textwidth}
  \includegraphics[width=\linewidth]{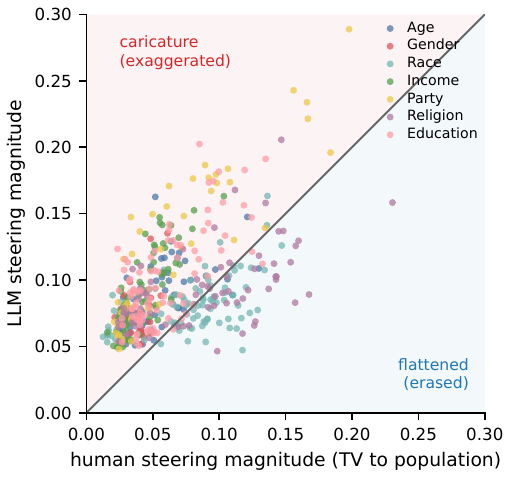}
  \caption{}
\end{subfigure}
\caption{\textbf{Single-feature steering: wrong hierarchy, compressed
magnitudes.} \textbf{a}, Mean steering magnitude (TV between subgroup and
population distribution, averaged over the wave's questions and the
dimension's values) by topic and dimension, for humans (left) and
GPT-4o-mini (right); \textbf{dots} mark each topic's strongest dimension;
unweighted sensitivity basis (Methods); right
strip, per-topic Spearman correlation between the two rankings. \textbf{b}, Simulated versus real steering magnitude, one
point per dimension-value $\times$ wave ($n=429$); 80\% of points lie
above the identity line, but race and religion---the strongest real
drivers---are the only dimensions not exaggerated.}
\label{fig:hierarchy}
\end{figure}

\section*{Results}

\subsection*{Single features: an inaccurate, compressed hierarchy}

Before moving to intersectional simulation, we first characterize how models treat each identity feature alone. Let us denote a demographic profile $g$ and question $q$, let $p_g$ be the real answer distribution of subgroup $g$ and $p_{\mathrm{pop}}$ that of the full population, with $\hat{p}_g$ and $\hat{p}_{\mathrm{pop}}$ their simulated counterparts (the model's output when conditioned on the profile and when prompted for an average American, respectively). From these we define (Fig.~\ref{fig:schematic}) the human \emph{steering} vector $s_g = p_g - p_{\mathrm{pop}}$ (how the subgroup's answer distribution differs from the population's) and its simulated counterpart $\hat{s}_g = \hat{p}_g - \hat{p}_{\mathrm{pop}}$; we summarise magnitudes by total variation (TV) distance throughout (the share of respondents who would have to change answers to match the comparison distribution).

Concretely, on a question about people's worry about paying bills (\emph{a lot}/\emph{a little}/\emph{not at all}), the weighted population answers $(0.34, 0.40, 0.26)$ and respondents earning \$100{,}000+ answer $(0.12, 0.43, 0.45)$: $s_g = (-0.22, +0.02, +0.20)$, $\tv =22$\% of the subgroup would need to change answers. The question throughout is whether $\hat{s}_g$ points where $s_g$ points and is as large. Both failure modes appear in this battery: GPT-4o-mini steers high earners the right way ($\cos = 0.81$) but too strongly - $2.5\times$, the real magnitude being ($\tv = 0.55$), while for Hispanic respondents on job worry---a real steer of $\tv = 0.23$---its simulated subgroup is identical to its simulated population, $\hat{s}_g = \mathbf{0}$.

Real opinion has topic-specific structure (Fig.~\ref{fig:hierarchy}a): religion is the strongest driver on most waves, race leads on questions of crime and the economy, party only on the political waves. The model hierarchy is topic-invariant: GPT-4o-mini ranks party first on 12 of 15 topics (mean rank correlation with the human hierarchy $\rho = 0.36$; the other models reach 0.46--0.55, still far from the human ordering, with the same party-first ranking). And steering magnitudes are compressed (Fig.~\ref{fig:hierarchy}b): across 429 dimension-value $\times$ wave points, the model exaggerates weakly-steering dimensions (gender: human mean TV $0.039 \rightarrow 0.074$; party: $0.071 \rightarrow 0.130$) while the two strongest real dimensions are the only ones \emph{not} exaggerated (race: $0.080 \rightarrow 0.080$; religion: $0.088 \rightarrow 0.093$). The model grants every dimension roughly the same distinctiveness; humans differentiate sharply.

\subsection*{Two features: collapse, not combination}

\begin{figure}[h!]
\centering
\includegraphics[width=0.95\textwidth]{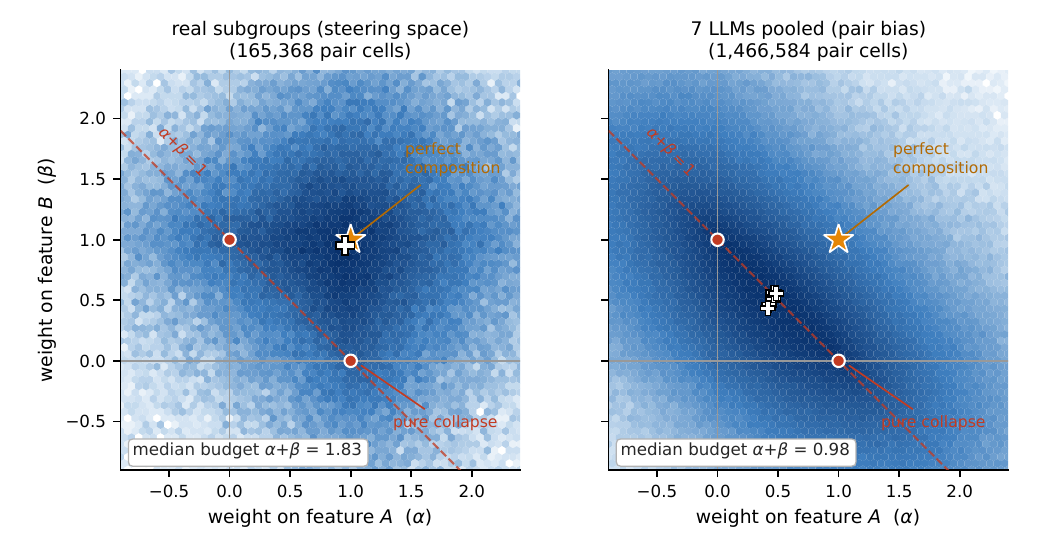}
\caption{\textbf{The identity budget: humans spend both features, every model spends one.} For each pair cell, the realized vector is decomposed by least squares into $\alpha$ times its $A$-feature part plus $\beta$ times its $B$-feature part; panels show the density of $(\alpha,\beta)$ across cells (log-scale hexbins; white cross is the median).
Perfect composition is $(\alpha,\beta)=(1,1)$, that is, full weight on both features (star). Pure collapse is $(1,0)$ or $(0,1)$, that is, full weight on one, none on the other. Left graph is ground truth (steering space): the density centers on perfect composition, median $(0.95, 0.95)$, with a median total weight $\alpha{+}\beta$ of 1.83---humans spend nearly two identities' worth of weight. Right graph is the model pair biases pooled over seven models (bias space; per-model panels in Extended Data Fig.~2): the density instead rides the line $\alpha{+}\beta=1$, a fixed budget of one identity's worth of weight (pooled median total 0.98; 0.91--1.03 per model). The budget is not split evenly between the two features: the median dominant share $\max(\alpha,\beta)/(\alpha{+}\beta)$ is 0.88, with 63\% of cells above 0.75, so the mass sits near the two collapse endpoints rather than midway. (The pooled coordinate-wise median, $(0.45, 0.50)$, is low only because \emph{which} feature dominates varies by cell; it does not indicate even attenuation. Similarly, the coordinate-wise medians on the left need not sum to the median total.) No model escapes the ridge.}
\label{fig:humancontest}
\end{figure}

The central question is whether a model prompted with two identities actually conditions on both. Its outputs alone cannot answer this, a pair simulation can sit anywhere in the response space, but its \emph{errors} can. Each single-feature bias $e_A = \hat{p}_A - p_A$ is a reproducible signature of how the model distorts that identity: which options it over- or under-weights whenever the feature enters a prompt. These signatures act as tracers. Because real subgroups compose additively (Fig.~\ref{fig:humancontest}), a model that integrated both conditionings would carry both distortions into the pair output at once, $e_{AB} \approx e_A + e_B$; a model that silently kept one feature would reproduce that feature's signature alone, $e_{AB} \approx e_A$, as if the second had never entered the prompt. The realized pair bias $e_{AB} = \hat{p}_{AB} - p_{AB}$ thus records which identities the model actually used. For every pair cell we therefore run a contest: the \emph{additive} prediction $e_A + e_B$ against the \emph{best single} feature's bias alone, each scored by cosine similarity to $e_{AB}$.

Figure~\ref{fig:schematic}b--c shows the two hypotheses geometrically: a composing model makes both single-feature mistakes at once, so its pair bias points along the sum; a collapsing model reproduces one feature's error as if the other had never entered the prompt. Per cell, both candidates are scored by cosine similarity to the realized $e_{AB}$; if conditioning composed, the additive prediction should win.

It loses, the best single dimension explains the pair bias better than the additive combination in \textbf{78.4\%} of profile--question cells for GPT-4o-mini (95\% CI [77.8, 79.0], cluster bootstrap over waves), with 75.3--81.2\% across the other seven models (Table~\ref{tab:models})---including 79.1\% for GPT-5.5 and 77.8\% for Claude Sonnet 5, two current-generation frontier models from different providers, each evaluated on the same two-wave subset.

A raw win rate, however, has no intrinsic chance level interpretation: the best single is a \emph{post-hoc} maximum over two candidates, and where the two single biases point much the same way, run-to-run noise decides the contest essentially at random---so the singles win well over half of near-parallel cells \emph{despite perfect composition}. The chance level cannot be derived a priori; it depends on the geometry of the model's own biases and the size of its own noise. We therefore measure the level by building synthetic pair cells in which the truth is
known by construction: \emph{additive-truth} cells, $e^{*}_{AB} = e_A + e_B + \eta$, assembled from the flagship's (GPT-4o-mini; our reference model throughout, Methods) real measured
single-feature biases, and \emph{collapse-truth} cells, $e^{*}_{AB} = e_{\mathrm{dom}} + \eta$, with $\eta$ drawn not from a model of noise but from the noise itself---deltas between four repeated end-to-end runs of the same wave, scaled by $1/\sqrt{2}$ and matched on option count ---and ran the identical contest, post-hoc maximum and all, on both sets. When the truth is exactly additive, the best-single still wins \textbf{40.3\%} of contests; when the truth is pure collapse, it wins \textbf{84.3\%}, not 100\% (noise sometimes tips the realized vector toward the sum). These two numbers are what the contest reports when each hypothesis is \emph{true}; an observed win rate is meaningful only relative to them. Rescaled between the endpoints (0 = additive, 1 = collapse), GPT-4o-mini's 78.4\% becomes a collapse index of $0.87 = (78.4-40.3)/(84.3-40.3)$ on the shared flagship endpoints; re-deriving the endpoints from each model's \emph{own} run-to-run noise---the primary, self-calibrated form---puts GPT-4o-mini at \textbf{0.86} and every model at \textbf{0.83--0.95} (Table~\ref{tab:models}, Index$_{\mathrm{self}}$; 95\% CI $\pm0.02$ on the flagship's index, Methods), about six-sevenths of the way to pure collapse. The scale is a property of the collapse, not of which model's noise floor sets it.

Absolute similarity values, by contrast, flatter the additive model and must be read against a null baseline (Extended Data Fig.~3a). The bias of a \emph{random unrelated} dimension achieves median cosine 0.84 with the pair bias; a high floor that is itself diagnostic, reflecting a shared population-level bias present in every conditioned output. Against that floor the additive prediction reaches 0.94 and the best single 0.96, and the additive prediction beats the random-dimension null in only 70\% of cells. Magnitudes complete the picture (Extended Data Fig.~3): the realized pair bias is barely six-tenths the size the additive model predicts (median ratio 0.59).

Nor is the collapse an artefact of small-cell ground-truth noise. Restricting to human pair cells with $n\ge200$---far stricter than the $n\ge20$ gate used throughout---the win rate edges down about a point (e.g.\ GPT-4o-mini $78.4\to77.3$\%). The small monotone decline is expected: best-single is a post-hoc maximum, upward-biased under target noise---inflation the endpoints already absorb---so cleaner cells shrink it. Depth 3 behaves the same way ($7.8\to8.6\%$ additive share at $n\ge200$): cleaner ground truth makes the model look marginally \emph{less} collapsed at both depths, exactly as the noise account predicts. The raw share also varies with option count---which is why the endpoints are option-count-matched---but the calibrated index is flat where the raw share slides (0.84 on $K{=}3$--5 items, 92\% of cells; 0.82 on $K{=}6$--8): the gradient is geometry, not substance (Methods). Given two identity features, models do not add them; they keep one and shrink or ignore the rest.

\subsection*{The collapse survives every elicitation format}

Every distribution so far comes from a single elicitation: the model distributes 1{,}000 hypothetical respondents across the options. A natural objection is that collapse is an artefact of that aggregate task rather than of demographic conditioning itself. We therefore re-ran the depth-2 contest under the two paradigms used elsewhere in the literature, changing nothing but the readout (Extended Data Fig.~4, Methods): \emph{individual sampling}, one simulated persona per call at temperature 1.0, histogrammed over $N{=}100$ personas per cell ($\approx$10 per cell--question pair); and \emph{log-probability readout}, a single forward pass softmaxed over option-letter logits with no sampling at all---the OpinionQA paradigm.

The collapse is unmoved, across five model families, the best-single win rate stays within \textbf{75.3--83.4\%} in every (model $\times$ paradigm) cell; per-model shifts from the aggregate value are small and run in both directions, and none escapes the band. Whether the model emits an aggregate count, role-plays one respondent at a time, or never samples at all, two identity features still collapse to one---a property of how the model conditions on demographics, not of the prompt used to read it out.

These formats vary the \emph{readout}; the collapse is equally indifferent to the \emph{conditioning} prompt, making it prompt invariant (Methods). Reframing the entire task---neutral and first-person system framings---moves Gemma-2-9B's and Claude Haiku 4.5's rates by $\le2$ points on identical cells. Chain of thought cannot buy composition either: explicitly directing the model to integrate both identities, or to reason step by step before answering, leaves the flagship unchanged and moves Claude Haiku 4.5 \emph{upward} (per-variant rates: Methods). Across two providers, two waves, and two scoring designs, the practitioner's direct remedy is inert.

\subsection*{Real subgroups compose additively---and grow more distinctive as identities intersect}

The contest presumes a benchmark worth validating: are real subgroups themselves additively structured? We ran the identical contest on the human side in steering space, for every pair cell with $n\ge100$. Humans do compose---but a perfect score here is not 100\%: the opponent is a post-hoc best single, advantageous to unitary bias if the target is observed through sampling noise and near-tie cells become coin tosses the maximum harvests. A \emph{perfectly additive} composer, simulated through multinomial noise at the same cell sizes, tops out at \textbf{49.3\%} ($n\ge100$; 52.5\% at $n\ge200$)---that ceiling is what perfection looks like under this rule, every win rate should be read as distance from it. Real subgroups score 44.4\% of 236{,}752 cells ($n\ge100$), rising to 47.8\% at $n\ge200$: \textbf{five points} short of perfect composition. GPT-4o-mini, scored on the identical cells in its own steering space, lands at 9.3\%---against its \emph{own} measured ceiling of 29.7\%, \textbf{less than a third} of what perfect composition would deliver, versus the humans' \textbf{ninety percent}. The human five-point gap behaves like noise, not failure: the win rate rises in lockstep with the ceiling as cleaner cells halve the noise, and the noise-immune counterpart---the coefficient median is indistinguishable from perfect composition.

This steering-space contest is the direct human comparison---both sides scored by the same rule in the same space---and it is \emph{stricter} on the model than the bias-space figure of Table~\ref{tab:models} (where additive wins $\sim$22\%), because the bias-space additive predictor carries the population-bias term twice (Methods, e\textsubscript{pop} correction). The 44.4\% versus 9.3\% on identical cells---90\% versus 31\% of each side's own attainable ceiling---is the cleanest single statement of the gap (the model's 29.7\% ceiling is measured from its own steering vectors and run-to-run noise, below the human 49.3\% because its noise is larger relative to its steering; construction in Methods).
Figure~\ref{fig:humancontest} shows the same fact without a contest: human subgroups centre on full weight for \emph{both} features (median $(\alpha,\beta)=(0.95,0.95)$, total 1.83), while every model's density rides the $\alpha{+}\beta{=}1$ line---one identity's worth of weight (median total 0.98--1.02), lopsided rather than split (the flagship's median dominant share is 0.92). The additive benchmark is not aspirational; it is very close to what real populations do, and simulations do not.

This contest tests the \emph{direction} of human composition; the magnitude composes too. Across the same pair cells, the noise-corrected excess in squared distinctiveness of the realised pair steering over the additive prediction $s_A+s_B$ is $-0.0008$ (95\% wave-cluster CI $[-0.0011, -0.0006]$): essentially zero, and if anything very slightly \emph{sub}-additive. Put plainly, stacking two identities moves a real subgroup's opinions exactly as far as the two tilts summed would predict---no further. Real pair subgroups thus carry no super-additive (emergent) surplus, in direction or in magnitude. This matters for how the models' failure should be read. Had real intersections carried emergent structure beyond their single-feature tilts, the additive benchmark would overreach, and a model's failure to match it could be excused as failure at a task real populations do not perform either. The data foreclose that defense: the target the models miss is plain addition---the easiest possible form of composition---and real subgroups hit it almost exactly. The models fail the easy version of the task.

Beyond serving as that benchmark, \textbf{this is a substantive result in its own right}: real intersectional subgroups are well approximated as the additive combination of their single-feature tilts in \emph{both} direction and magnitude, yet---corrected for sampling noise---grow $2.5\times$ more distinctive in squared distance from one to three features (Extended Data Fig.~1); the two facts are not in tension, since adding non-orthogonal tilts moves a subgroup steadily away from the population mean. This near-additivity converges with the MAIHDA literature's finding that additive main effects dominate between-stratum variance\citep{evans2018multilevel,merlo2018multilevel}, and is compatible with deep interactions mattering for \emph{individual-level} prediction\citep{ghitza2013deep}: marginals can be near-additive while individual outcomes are not, and marginals are the level at which silicon sampling operates. To our knowledge this joint characterisation---additive yet increasingly distinctive---has not been quantified at this scale. It is a claim about one country, one language, and one instrument family (US English-language ATP closed-form items, 2017--2021); portability elsewhere is untested.

\subsection*{The kept dimension is chosen nearly blindly and suppresses race and religion}

\begin{figure}[h!]
\centering
\begin{subfigure}{0.55\textwidth}
  \includegraphics[width=\linewidth]{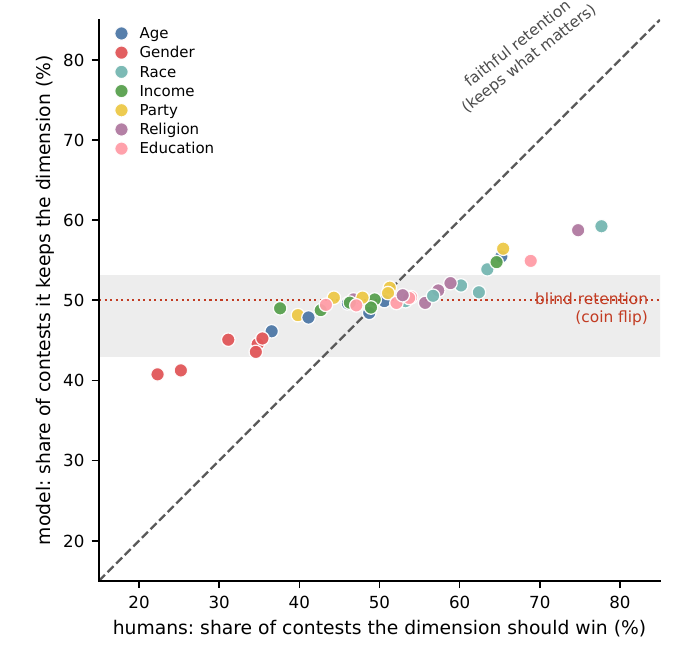}
  \caption{}
\end{subfigure}\hfill
\begin{subfigure}{0.44\textwidth}
  \includegraphics[width=\linewidth]{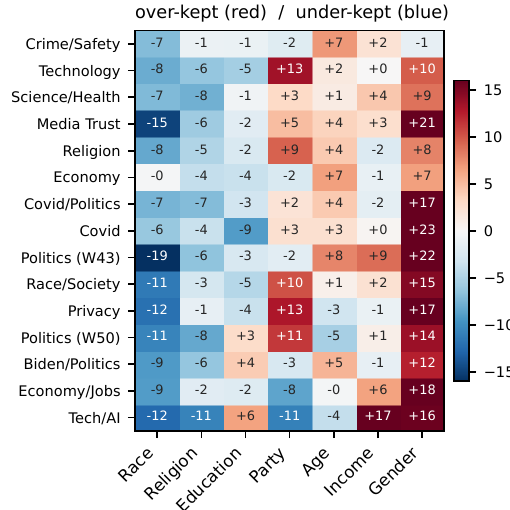}
  \caption{}
\end{subfigure}
\caption{\textbf{Near-blind, topic-invariant collapse direction.}
All panels are computed in bias space. \textbf{a}, Each point is one
(dimension, partner) pairing, pooled over the five full-coverage models:
how often humans say the dimension should win ($x$) against how often the
model keeps it ($y$). Faithful retention is the diagonal; the data lie
nearly flat inside the coin-flip band (shaded, 43--53\%) while human
relevance spans 22-78\%. \textbf{Retention is near-blind to relevance} (full pairwise dominance matrix: Extended Data Fig.~6).\textbf{b}, Keep-rate minus human-dominance rate (percentage points) by topic: the signs barely move across topics---race is under-retained on all 15 waves, including the racial issues focused wave.}
\label{fig:direction}
\end{figure}

Which feature survives? For every collapsed pair cell we compared the \emph{winner} (the dimension whose single bias best matches the pair bias) with the \emph{human-dominant} dimension (whose real subgroup deviates farther, in TV, from the population). Agreement is 53.3--57.9\% across all eight models---and because dominance rates differ by dimension, chance is not 50\% but 50.4--52.2\% (permutation null), so models exceed genuine chance by only \textbf{3--7 points}: weakly relevance-tracking, dominated by a fixed prior. Nearly all the excess is carried by party (retained 54--62\% when human-dominant versus 42--47\% when not), the only dimension whose retention tracks relevance at all.

The residual preferences form a fixed, topic-invariant prior (Fig.~\ref{fig:direction}). Models over-retain the dimensions that drive real opinion least (gender: kept 12.5 points more often than humans warrant, pooled over the five full-coverage models; income $+3.0$) and under-retain the dimensions that drive it most in the ground truth (race $-9.5$; religion $-5.0$); race is significantly under-retained in \emph{all six} prior-generation models with race-pair coverage ($-8.3$ to $-10.6$ points, every bootstrap CI excluding zero), on all 15 topics---including the wave about racial issues ($-11.4$). The specific tilt, however, is slightly generation-dependent: on matched waves and cells, every tilt is modestly attenuated in GPT-5.5, while Claude Sonnet 5, the second 2026 model, shows no attenuation at all---its race suppression is at full prior-generation strength---so whatever softened GPT-5.5's tilt is specific to that model, not to the generation (matched-wave deltas: Methods). The compression toward a coin flip is unchanged (keep-rates 41--53\% on the matched cells; agreement 56.2\%). The structural failure persists across generations; the direction of its tilt tracks whatever the training regime encodes.

Within the discarded dimensions, moreover, it is the minority identities that are dropped: the flagship's value-level keep-minus-dominance deltas run $-19$ to $-23$ points for Black, Hispanic, Asian, mixed-race, atheist and Jewish respondents, against $+16$ for White and $+4$ to $+6$ for Protestant and Catholic (Methods). Aggregate ``race suppression'' is, concretely, the discarding of minority racial and religious identities while the majority value is over-retained.

The prior has a structure that ``party override'' narratives miss\citep{santurkar2023whose}. In the model's own expressive space (which single feature's \emph{steering} a pair's steering resembles; no ground truth involved), party defeats race (73\%) and religion (72\%) yet \emph{loses} to age, education, and income; overall win rates run age 63\% down to race \textbf{39\%} and religion \textbf{35\%}, which lose to every other dimension when combined. What looks like party dominance is a special case of a broader pattern---\textbf{sensitive-dimension suppression}: models differentiate loudly on party while writing race and religion out of any combination, a permission structure already present upstream of preference-tuning (below) and so reflecting corpus-level norms rather than alignment policy.

\textbf{Most of this suppression predates preference-tuning.} In a three-way contrast on Llama-3.1-8B---base weights under a base prompt, instruct weights under the same prompt, instruct weights under the chat template---the race deficit is already $-10.8$ points in the \emph{base} model and barely moves: $+0.6$ from tuning the weights ($95\%$ CI $[-3.5,+4.7]$), $0.0$ from the chat template. Religion behaves identically ($-5.3$ at base; shifts $-0.6$, $-0.4$). These nulls are informative: the design is powered to detect shifts of $5.9$ (race) and $2.7$ (religion) points---over half the suppression itself---so post-training neither installs the deficit nor removes more than that bound. Mistral-7B replicates the results (base deficit: race $-7.3$, religion $-3.6$; tuning shifts $\Delta=-1.2$ and $+1.9$ against MDEs of $5.8$/$5.3$). Two decomposable architectures thus locate the suppression upstream of preference-tuning; we make no claim about the closed models. These are precisely the two dimensions real opinion runs on.

\subsection*{Three features: an identity budget}

Extending the contest to three-feature profiles shows the collapse is a capacity limit, not a peculiarity of pairs (Extended Data Fig.~5). The fully additive predictor wins in only \textbf{7.8--10.0\%} of cells across the seven models with triple coverage (per-model shares and CIs, Table~\ref{tab:models}). The best sub-pair instead explains 52--66\% of cells and the best single roughly a third: in at least 90\% of cases a three-feature persona behaves as at most two of its identities. Models operate within an \emph{identity budget} of one to two dimensions, discarding the rest (machinery validated by the human control, which recovers $(0.95,0.95)$, not the ridge; Fig.~\ref{fig:humancontest} and Methods).

The calibration sharpens this: on synthetic triples, a perfectly additive model would produce an additive share of 67.2\%, while a pure collapser---whose additive predictor still beats the post-hoc maximum of six rivals occasionally by chance---concedes 6.6\%. Re-deriving the endpoints from each model's \emph{own} noise puts every model's depth-3 index at \textbf{0.92--0.98} (Table~\ref{tab:models}): the observed shares sit essentially on the collapse floor.

\subsection*{Added identity information buys no accuracy}
Figure~\ref{fig:spine} summarizes prediction error across depth, models, and the human sampling-noise floor, with four consistent patterns. First, every model sits far above the irreducible error at every depth. The ground truth is itself an estimate: a cell of $n$ respondents yields a noisy version of its "true" distribution, so even a perfect simulator would score a nonzero TV against it. We measure this floor empirically by splitting each real cell into two random halves and computing the TV between them---two estimates of the same population, so their distance is pure sampling noise: 0.079, 0.120, and 0.147 at depths 1, 2, and 3 (matching multinomial theory to within 2\%; conservative---Methods). Models exceed this floor by 1.8--2.7$\times$ (GPT-4o-mini) up to ${\sim}4\times$ (the weakest). Nor does the forced-choice refusal artefact close the gap: removing its bound (5.1\% simulated refusal mass versus 0.7\% real; Methods) only lowers GPT-4o-mini's depth-1 error from 0.212 to 0.168---still twice the 0.079 floor.

\begin{figure}[h!]
\centering
\includegraphics[width=0.62\textwidth]{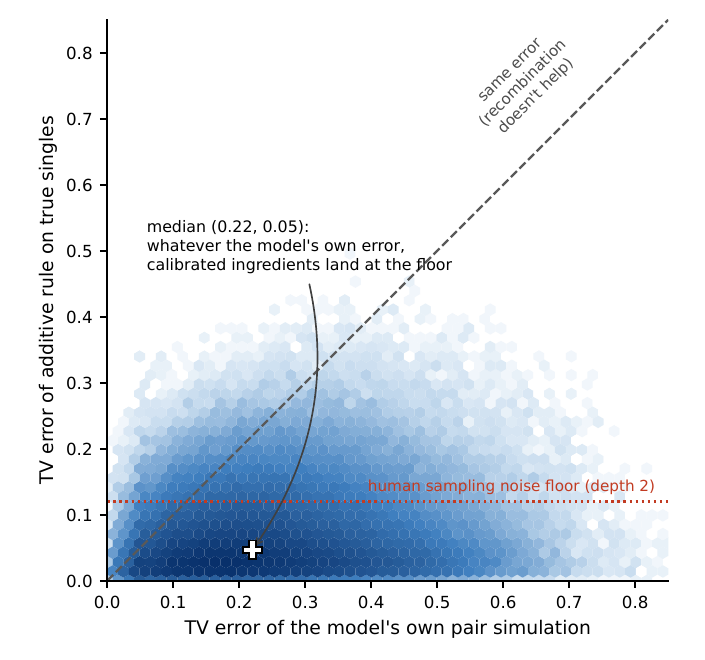}
\caption{\textbf{The accuracy gap is single-feature miscalibration, not
the composition rule.} The panel asks whether the additive rule itself
is capable of accuracy, by feeding it perfect ingredients. Each hexbin
aggregates flagship pair cells placed by two errors: $x$, the TV error
of the model's own pair simulation---how far its native two-feature
output sits from the real subgroup; $y$, the TV error of the additive
rule applied to the \emph{true} single-subgroup distributions from the
survey, $\max(p_A{+}p_B{-}p_{\mathrm{pop}},0)$, renormalised, meaning, how far
simple addition lands from the same target when its inputs carry no
model error. If accuracy were limited by the composition rule, $y$
would track $x$ (dashed identity line: recombination doesn't help).
Instead the cloud is flat and low: whatever the model's own error---the
$x$-axis spans 0 to 0.8---addition with calibrated ingredients lands at
the human sampling-noise floor (dotted; medians $0.22 \rightarrow
0.05$, white cross). The additive rule is sufficient; the error lives
in the model's single-feature outputs, which are miscalibrated before
they are ever combined. Consistently, applying the same rule to the
model's \emph{own} singles is marginally \emph{worse} than its native
pair output (per-model means, Extended Data Fig.~7). $N = 336{,}042$
cells.}
\label{fig:decomp}
\end{figure}

\begin{figure}[tp]
\centering
\includegraphics[width=0.62\textwidth]{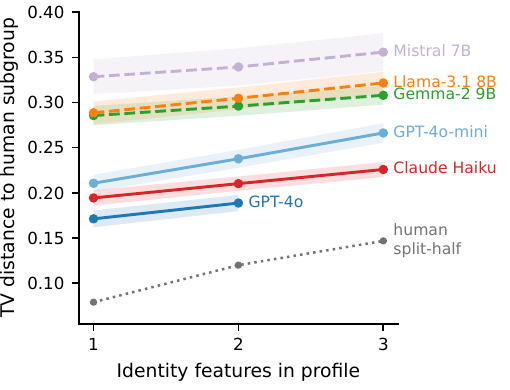}
\caption{\textbf{Simulation error by profile depth, against the human sampling floor.} Mean TV between simulated and real subgroup distributions, by number of identity features; bands, 95\% CIs over waves; grey, split-half TV between random halves of the same human cells ($n\ge 40$). Model error sits above the floor at every depth and is flat in depth once the rising floor is accounted for.}
\label{fig:spine}
\end{figure}

Second, depth has remarkably little effect on model error: the noise floor rises with depth while model error stays within the floor-adjusted band (GPT-4o-mini 0.212/0.238/0.269; TV does not subtract linearly, so read as bounds). Additional identity dimensions contribute little information---precisely what collapse predicts.

Third, model ranking is largely determined by overall capability rather than by the ability to compose identities. Larger and more capable models achieve uniformly lower prediction error---GPT-4o ranges from 0.171 to 0.188, whereas Mistral-7B ranges from 0.329 to 0.356---yet all exhibit essentially the same dependence on intersectional depth: scale improves baseline calibration, not composition. A limited two-wave probe (W26, W34) suggests the pattern may persist across model generations. GPT-5.5 reduces prediction error by 34--38\% relative to GPT-4o-mini on the same data, yet their collapse statistics remain indistinguishable (79.1\% versus 78--80\%). The constituent identities are estimated more accurately, but they are discarded at essentially the same rate.

Finally, the behavior of real populations contrasts sharply with that of the models. After correcting for sampling noise, genuine human subgroups become progressively more distinctive as additional identities intersect. Unbiased squared distinctiveness increases from 0.0082 at depth 1 to 0.0147 at depth 2 and 0.0204 at depth 3, representing a 2.5-fold increase (Extended Data Fig.~1). Real intersectionality therefore produces increasingly distinctive response distributions, whereas simulated respondents do not.

Would composing fix it? We built the additive estimator from each model's own single-feature predictions ($\hat p_A + \hat p_B - \hat p_{\mathrm{pop}}$; depth 3 analogously) and scored it against the real subgroups. It provides essentially no improvement---slightly \emph{worse} at depth 2 (e.g.\ GPT-4o-mini $0.238\to0.247$ TV; the estimator beats the native output in under 45\% of cells at both depths). But the identical rule applied to the \emph{true} single-feature distributions ($\max(p_A + p_B - p_{\mathrm{pop}},0)$, renormalised) reaches a mean depth-2 TV of 0.062 for GPT-4o-mini (per-cell medians $0.22\to0.05$, Fig.~\ref{fig:decomp})---below the human noise floor and a quarter of the model's own 0.238. Nearly the entire accuracy gap is therefore single-feature miscalibration, not the composition rule. This sharpens rather than softens the collapse result: \textbf{composition failure and calibration failure are separate defects that compound---perfect additive composition of the model's existing singles would not restore accuracy, and the model demonstrably does not compose in the first place.}

Model unreliability further compounds the problem: across four repeated end-to-end runs, run-to-run TV is $\approx$0.119 at \emph{every} depth---above the human floor at depth 1, comparable at depths 2--3, and indifferent to how much identity information the prompt contains (Extended Data Table~1). The instability is confined to individual cells: the across-run spread of the aggregate errors in Fig.~\ref{fig:spine} is below 0.001 TV, so prompt stochasticity explains no reported effect.

\begin{figure}[h!]
\centering
\includegraphics[width=0.95\textwidth]{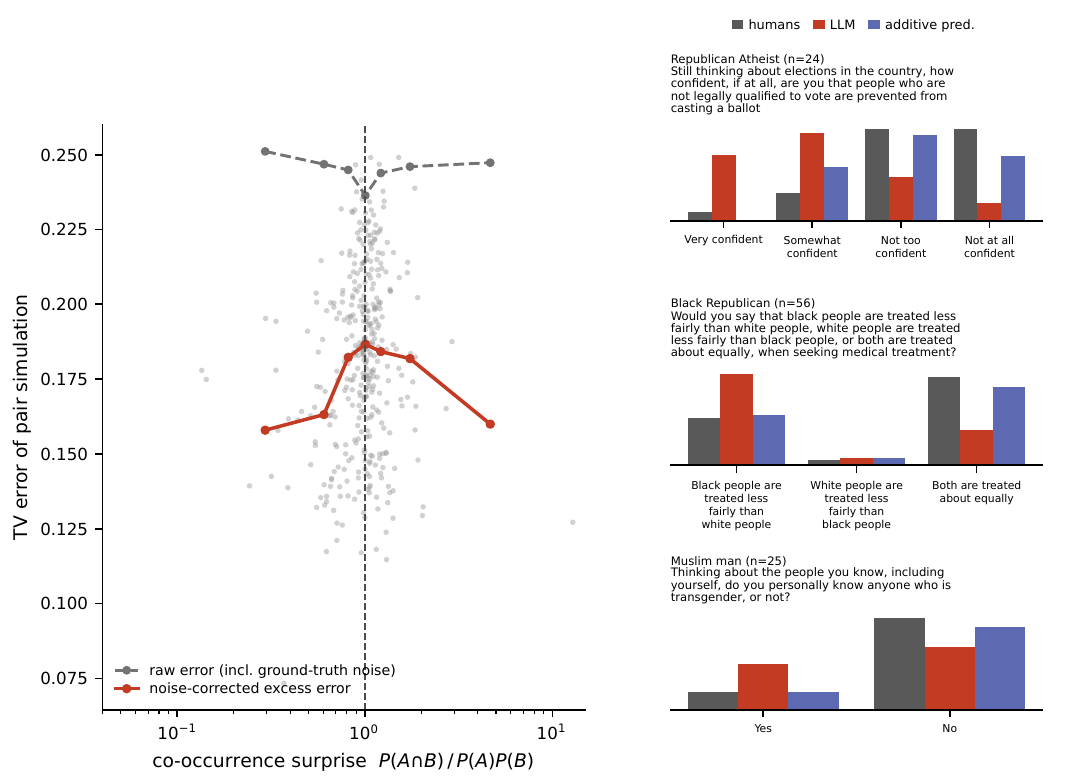}
\caption{\textbf{Failure is uniform across profile typicality.} Left, mean simulation error by co-occurrence surprise (binned; log scale; $\le$1 = counter-stereotypical). Raw error (grey) is highest in the rarest bin, but that elevation is ground-truth sampling noise from small cells: the Republican-atheist profile's raw 0.277 decomposes into 0.178 genuine error—below the average typical profile—plus 0.099 noise. Noise-corrected excess error (red) shows no counter-stereotypical penalty—the rarest bins sit, if anything, slightly below the typical ones. Right, case studies (one question each, selected by a fixed rule to preclude curation; human cell size on each panel; additive composition wins in 13–32\% of these profiles' cells and, aggregated over all cells, does not beat the native output): the additive prediction (blue) tracks the real distribution (grey) better than the model's pair simulation (red).}
\label{fig:uniform}
\end{figure}

\subsection*{Failure is uniform and not conditioned on specific subgroups}

Prior work predicts that incongruous personas---those rare in the real population---should be hardest to simulate\citep{liu2024evaluating,cheng2023compost}. Raw data appear to agree: profiles with the lowest co-occurrence surprise
$P(A{\cap}B)/P(A)P(B)$ (Republican atheists, Black Republicans; surprise 0.14) show the highest raw error (Fig.~\ref{fig:uniform}). But raw error also correlates with ground-truth cell size ($\rho=-0.26$): rare profiles have small human cells, and small cells have noisy ground truth. Subtracting each cell's expected sampling noise leaves a weak \emph{positive} residual (surprise versus excess error: Spearman $\rho=+0.13$, two-sided $P=0.01$---if anything, \emph{typical} profiles are marginally worse); restricting to large cells ($n\ge100$), where no correction is needed, the raw relationship is flat ($\rho=+0.04$, n.s.). Both analyses agree: the predicted counter-stereotypical penalty is absent, and any residual tilt runs the opposite way.

Simulation failure is therefore \textbf{uniform across profile typicality}: no typicality-defined safe subset, and no especially dangerous one---the signature of a mechanism that discards features blindly. The methodological warning: audits that do not control ground-truth cell size will rediscover a spurious counter-stereotypical effect, as our own raw data did before correction; prior reports\citep{liu2024evaluating,cheng2023compost} measure open-ended text on precisely the rare profiles where uncorrected noise is largest. The case studies in Fig.~\ref{fig:uniform} isolate the mechanism: in each, the additive prediction from the model's own singles tracks the real distribution better than its actual pair simulation---the model had the ingredients and did not combine them.

Simulations also distort within-group agreement: simulated distributions are \emph{more} dispersed than real ones at every depth ($\Delta H = +0.18$--$0.23$ nats, estimator-robust; Extended Data Table~3, Methods)---the much-discussed ``flattening'' of identity groups\citep{wang2025large} is \emph{between}-group, not within: a hedge, not a caricature.

\begin{table}[h!]
\centering
\small
\setlength{\tabcolsep}{4.5pt}
\begin{tabular}{lcccccc}
\toprule
Model & Depths & Rows & Collapse (d2) & Index$_{\mathrm{self}}$ & Additive share (d3) & Index$_{\mathrm{d3}}$\\
\midrule
GPT-5.5$^{\dagger}$         & 1--3 & 0.19M & 79.1\% [78.3, 79.9]$^{\ddagger}$ & 0.90 & 8.9\% [7.9, 10.1]$^{\ddagger\S}$ & 0.94\\
Claude Sonnet 5$^{\dagger}$ & 1--3 & 0.20M & 77.8\% [77.0, 78.7]$^{\ddagger}$ & 0.83 & 8.6\% [7.4, 10.0]$^{\ddagger\S}$ & 0.92\\
GPT-4o                      & 1--2 & 0.86M & 77.6\% [77.0, 78.2] & 0.86 & -- & --\\
Claude Haiku 4.5            & 1--3 & 3.70M & 77.9\% [77.3, 78.4] & 0.85 & 8.6\% [8.3, 9.0] & 0.94\\
GPT-4o-mini                 & 1--3 & 2.76M & 78.4\% [77.8, 79.0] & 0.86 & 7.8\% [7.4, 8.2] & 0.98\\
Mistral-7B                  & 1--3 & 2.74M & 81.2\% [80.5, 82.0] & 0.90 & 8.1\% [7.8, 8.5] & 0.97\\
Gemma-2-9B                  & 1--3 & 2.53M & 75.3\% [74.6, 76.1] & 0.84 & 9.2\% [8.9, 9.6] & 0.94\\
Llama-3.1-8B                & 1--3 & 2.74M & 80.3\% [79.8, 80.8] & 0.95 & 10.0\% [9.7, 10.4] & 0.94\\
\bottomrule
\end{tabular}
\caption{\textbf{Models, coverage, and headline replications.} Rows:
successful simulated distributions collected for the model, including
its W26 repeat runs.
Collapse (d2): share of pair cells where the best single feature explains
the bias better than the additive combination.
Index$_{\mathrm{self}}$: that share rescaled between calibration
endpoints derived from the model's \emph{own} run-to-run noise, so 0 = perfect composition and 1 =
pure collapse; a shared-endpoint variant calibrated on the flagship's
noise tracks these values within 0.04 (Methods); under the population-bias--corrected additive predictor ($e_A{+}e_B{-}e_{\mathrm{pop}}$, the fair additivity test in bias space; Methods), the indices span 0.80--0.95, so conclusions are stable under both forms. Additive share (d3):
share of triple cells where the full additive combination wins.
Index$_{\mathrm{d3}}$: that share rescaled between depth-3 calibration
endpoints derived per model from its own run-to-run noise (Methods);
Brackets, 95\% cluster-bootstrap CIs.
$^{\dagger}$Current-generation
robustness checks, two-wave subset (W26, W34); the two waves agree
closely.}
\label{tab:models}
\end{table}

\section*{Discussion}

Eight models from five organizations, spanning multiple orders of magnitude in scale and two model generations, produce the same numbers. At two features, the best single feature explains the pair bias in 75--82\% of cells---against measured endpoints, six-sevenths of the way to pure collapse. At three features, the full additive combination wins only 7--10\% of contests, essentially the pure-collapse floor (Index$_{\mathrm{d3}}$ 0.92--0.98). Which feature is retained tracks true relevance only 3--7 points above a permutation null---near-blind. Prediction error is flat as identities are added, where genuine integration would use the extra information. And failure is uniform across profile typicality: no census statistic flags which personas to distrust. The same collapse appears under entirely different elicitations---individual-persona sampling and log-probability readout (75--84\% across five families and three paradigms; Extended Data Fig.~4)---so it is a property of demographic conditioning, not of the aggregate prompt. This stability---including the two current-generation models, which collapse at the prior-generation rate (79.1\% and 77.8\%)---marks the limitation as structural, a property of how demographic prompting interacts with current training rather than a capability gap the next release will close.

The sensitive-dimension suppression has a definite locus, and it is not the obvious one. Instruction tuning is the natural suspect, but the base-versus-tuned contrast rules out a post-training origin for most of it: the race and religion deficits are at full strength in the base model under a base prompt, and neither tuning the weights nor switching to the chat template shifts them by more than a point, against a design powered to detect $2.7$--$5.9$-point shifts. Since the deficit predates post-training in the two decomposable families, the most economical reading is a provider-specific corpus shift rather than alignment policy or generation. But GPT-5.5's base weights are inaccessible, and one cross-generation contrast cannot separate scale, data, and provider practice. For the two open families, the collapse's \emph{form} is invariant and its \emph{direction} is set upstream of post-training; we do not extend that claim to the closed models.

Training-data contamination deserves treatment as a candidate \emph{mechanism}, not only as an accuracy confound. OpinionQA publishes single-feature subgroup distributions, and no intersectional tables of this kind exist in public corpora---so a model that retrieved memorised marginals and echoed the best-matching one would reproduce exactly the behavioural signature we call collapse. Four observations rule out retrieval as the mechanism. First, the single-feature outputs themselves sit far above the sampling floor (1.8--4$\times$), incompatible with verbatim retrieval of published tables. Second, retrieval cannot generate the \emph{directional} structure: the systematic suppression of race and religion relative to measured human dominance is a property of how features compete inside a combined prompt, not of any published marginal. Third, the suppression appears at full strength in base checkpoints under a document-completion prompt that shares no surface form with OpinionQA. Fourth---and most directly---retrieval predicts a gradient the data lack: cells whose parent singles best match ground truth (the plausibly memorised ones) should collapse most, yet stratifying the flagship's 335{,}979 pair contests by parent-single accuracy leaves the share flat (78.5/78.0/78.7\% across terciles). Memorisation, where present, biases accuracy \emph{toward} the ground truth (Methods); it cannot manufacture the compositional failure.

The two claims---invariant \emph{structure}, generation-dependent \emph{direction}---are not in tension: whether the model integrates a second feature is a structural property of the conditioning, while which feature it keeps is a calibration property of the individual biases; a newer model can recalibrate the tilt while still keeping one feature. 

For practitioners the implications are concrete, scoped to \emph{demographic-feature} conditioning and the US English-language ATP context (2017--2021); richer backstory- or interview-based conditioning\citep{argyle2023out,park2024generative} could in principle compose differently, though prompt reframing and chain-of-thought instruction did not move our findings. Within that scope, the marginal identity feature is informationally inert: enriching a persona from one attribute to three adds tokens, not information, and the model's test--retest noise (0.119 TV, depth-invariant) matches or exceeds the sampling noise of the real populations being replaced. Two failures should be kept apart: the \emph{accuracy} failure is dominated by single-feature miscalibration (the true-singles recombination reaches 0.062), so under current calibration collapse costs approximately nothing in accuracy; the \emph{structural} failure is what remains after miscalibration is fixed, and its significance is representational (whose feature is erased: disproportionately race and religion) and counterfactual (with calibrated ingredients, addition reaches the noise floor). The remedy for suppression, conversely, is calibration to \emph{measured} group differences, not maximal differentiation---leaning harder into race without ground truth reproduces the caricature failure mode\citep{cheng2023compost,wang2025large}. And the economic case inverts: silicon sampling is pitched as a cheap oversample of rare intersectional populations\citep{argyle2023out,sun2024random}---exactly the populations that require the composition current models do not perform. Meanwhile the phenomenon the method would erase is measurable: real subgroups grow $2.5\times$ more distinctive from one to three features. Intersectional distinctiveness is in the data\citep{crenshaw1989demarginalizing}. It is not in the simulations.

The constructive alternative exists: multilevel regression and post-stratification remains the standard for small-cell estimation, and our human near-additivity result is evidence for the main-effects-dominant structure it assumes---silicon sampling should be benchmarked against that, not against nothing. Two lessons extend beyond this study: similarity metrics between bias vectors are uninterpretable without a null floor (a random unrelated dimension scores 0.84 here), and ground-truth sampling noise is an active confound that manufactures findings. We recommend noise floors, null calibrations, and split-sample confirmation as defaults\citep{bisbee2024synthetic,dominguez2024questioning, beck2024sensitivity}. Our characterisation is behavioural; representation-level probing of whether the discarded feature is never encoded or encoded-but-down-weighted is the natural next step, which we leave open.

\section*{Methods}

\subsection*{Survey data and intersectional ground truth}
We use 15 waves of the Pew American Trends Panel (ATP; W26--W92, fielded 2017--2021), with 67--128 closed-form questions and 2{,}524--10{,}221 respondents per wave, covering crime, technology, science, media trust, religion, the economy, COVID-19, politics, race, and AI. From respondent-level micro-data, we computed the response distribution $p_g(q)$ of every demographic cell $g$ with $n\ge 20$ valid respondents: all single values, all pairs of values from distinct dimensions, and all triples, over seven dimensions (age: 4 values; gender: 2; race: 6; income: 5; political party: 3; religion: 5; education: 6; ``Refused'' is excluded as a demographic value). This yields 31 single-feature, 405 pair, and 1{,}355 triple profiles---1{,}791 distinct profiles per question at full coverage. All primary ground-truth distributions are survey-weighted using the ATP per-wave weights released by Pew (cell inclusion still gated on $n\ge 20$ raw respondents); the headline pipeline---spine, collapse shares, collapse direction, noise floor, bootstrap and permutation inference, split-sample confirmation---uses weighted distributions throughout. Unweighted distributions differ by mean TV 0.029, 0.040 and 0.053 for population, single and pair cells respectively---4--6$\times$ smaller than any model's error---and every headline number replicates under both treatments (Extended Data Table~2); the remaining secondary analyses (single-feature hierarchy, typicality, and entropy) use unweighted distributions and are bounded by the same sensitivity. Two temporal caveats apply to the \emph{accuracy} statistics (not to the composition contests, which compare the model's outputs to its own single-feature outputs and are internal to the model's geometry). First, the prompt does not anchor the response period, so part of the absolute error may reflect genuine opinion drift between fielding (2017--2021) and the models' training cutoffs (2024--2026) rather than misalignment. Second, ATP toplines and the derived OpinionQA release are public and plausibly present in pretraining corpora; memorisation would bias simulations \emph{toward} the ground truth, making the reported accuracy failures conservative.

\subsection*{Simulation protocol}
For each profile and question, the model is instructed (system prompt: expert demographic researcher; user prompt: fixed template) to distribute exactly 1{,}000 hypothetical respondents matching the profile across the answer options and output only the integer counts, which we normalise to $\hat{p}_g(q)$. Models tend to emit round-number counts (multiples of 25 or 50); this quantisation is part of the run-to-run variability that the measured $\eta$ absorbs, so it inflates no contest statistic. Eight models were run: GPT-4o, GPT-4o-mini, GPT-5.5, Claude Haiku 4.5, Claude Sonnet 5, Gemma-2-9B, Mistral-7B, and Llama-3.1-8B; sampling temperature 0.7 and a single question per call throughout, except GPT-5.5 and Claude Sonnet 5, which do not accept a sampling-temperature parameter and ran at their fixed provider defaults.
 
Features within a profile are listed in a fixed canonical dimension order---an arbitrary convention---so we verified that the retained dimension tracks the feature itself and not its position in the prompt: rerunning the depth-2 contest on Gemma-2-9B (W26) with the feature order \emph{reversed} leaves the retained dimension unchanged in 76\% of matched cells, and the slot-1 feature is retained 48\% of the time under canonical order versus 52\% under reversed (708 cells)---i.e.\ position carries essentially no signal, where a position-driven rule would retain slot-1 in both orders. The same control on Claude Haiku 4.5 is stronger still: the retained dimension is unchanged in 88\% of 767 matched cells, with slot-1 retention 46.9\% canonical versus 54.1\% reversed. The remaining flips are near-tie cells, not a systematic slot effect; the collapse-direction findings therefore reflect dimension identity, not presentation order. The framing control compares the pipeline's expert-researcher system framing against a neutral ``survey response simulator'' and a first-person ``simulate individual people'' wording; the instruction and chain-of-thought controls re-run the paired elicitation with an explicit joint-integration instruction or a reason-step-by-step preamble, with single-feature predictors re-elicited under the same instruction for the within-regime scoring (verbatim prompts in the released code). Per-variant depth-2 best-single rates: Gemma-2-9B 74.9/74.4/72.8\% and Claude Haiku 4.5 79.5/78.1/79.8\% across expert/neutral/first-person framings on identical cells; flagship 79.4\% baseline, 79.5\% instructed, 79.0\% chain-of-thought (79.9/80.2\% with singles re-elicited under the same instruction, ruling out a regime-mismatch artefact); Claude Haiku 4.5 79.7\% baseline, 83.1\% instructed, 85.5\% chain-of-thought. Both prompt-level controls (framing and feature order) are single-wave design checks run on two models---one open-weight (Gemma-2-9B) and one closed API model (Claude Haiku 4.5)---that replicate each other: they rule out the specific artefacts they target rather than certifying every possible prompt. Response options are presented in the instrument's canonical order throughout; option-order sensitivity is not separately tested.
 
GPT-4o-mini, and Claude Haiku 4.5 serve as the flagships because full depth-1--3 coverage at this scale (2.8M distributions per model) is cost-prohibitive for the larger hosted models; every \emph{singular} use of ``the flagship''---the shared calibration endpoints (40.3/84.3), the steering-space 9.3\%, the depth-3 index 0.98, the per-cell decompositions, and the reliability tables---refers to GPT-4o-mini, with Claude Haiku 4.5 as the second full-coverage replication; the open-weight models replicate every headline at full coverage (per-model replications: Extended Data Fig.~2). We pin exact snapshots for the hosted models where dated snapshots exist: \texttt{gpt-4o-mini-2024-07-18}, \texttt{gpt-5.5-2026-04-23}, and \texttt{claude-haiku-4-5-20251001}. Two hosted models could not be pinned, and we flag both as reproducibility caveats: GPT-4o was queried through the unversioned \texttt{gpt-4o} alias (accessed 2026-04 to 2026-06), and Claude Sonnet 5 was released during our collection window without a dated snapshot and was queried through its unversioned alias. Neither carries a load-bearing result: GPT-4o is corroborative only (depths 1--2; Table~\ref{tab:models}), Claude Sonnet 5 seconds GPT-5.5 on the two-wave current-generation subset, and the load-bearing depth-1--3 results rest on the pinned flagships and the open-weight models, so alias drift does not affect a headline number.
 
Triple-feature profiles cover all 20 three-way combinations of the six core dimensions (age, gender, race, income, political party, religion) for GPT-4o-mini, Claude Haiku 4.5, Gemma-2-9B, Llama-3.1-8B, and Mistral-7B.
 
GPT-5.5 serves as a current-generation robustness check on a two-wave subset (W26 and W34, all depths; 185{,}157 distributions); Claude Sonnet 5 mirrors GPT-5.5's exact cell set on the same two waves, giving the current generation a second provider. Their triple-feature profiles cover the four-dimension (age/gender/race/income) subset marked $^{\S}$ in Table~\ref{tab:models}---a cost-driven restriction; race- and religion-containing \emph{pairs} are fully covered for both models at depth 2, which is where the suppression contrast is drawn. These two waves are not the explicitly political waves of the panel---W26 concerns local crime, safety, and personal worry, W34 concerns science and research investment---so the GPT-5.5 collapse and its attenuated race deficit are not an artefact of party-aligned topic content (which would otherwise let Political Party dominate the conditioning).
 
To measure run-to-run reliability and to calibrate the collapse index per model, one wave (W26) was additionally run four times end-to-end for each of the eight models---for the flagship across profiles of every depth (2{,}964 single, 31{,}590 pair and 20{,}904 triple cells per run, the basis of the by-depth reliability in Extended Data Table~1), and for every model at minimum across single-feature profiles, supplying each model's own $\eta$---at temperature 0.7 matching the pipeline for every model that exposes the parameter, and at the fixed provider default for GPT-5.5 and Sonnet 5; the $\eta$-scaling sweep below bounds any effect of a residual settings mismatch on the calibration endpoints. The run-to-run distributional noise from these repeats supplies each model's own $\eta$ for its self-calibrated endpoints (Table~\ref{tab:models}, Index$_{\mathrm{self}}$). GPT-5.5's repeat set was truncated partway by a provider quota; its $\eta$ rests on the surviving repeat calls, a set still several times larger than the smallest open-weight repeat set.

Failed requests do not select on the sensitive dimensions: tallying the flagship's depth-2 records by involved dimension, race- and religion-containing pairs show failed-record rates at the low end (10.9--12.3\%, against 16.7\% of records overall), while the one elevated dimension---education, 41.9--42.8\%, uniform across its values---reflects an interrupted batch campaign rather than content-dependent refusals; failures are re-queued, and Table~\ref{tab:models} counts successful rows only, so no dimension-selective parse artefact enters the collapse-direction analyses.

\subsection*{Alternative elicitation paradigms}
To test whether collapse depends on the aggregate elicitation, we reran the depth-2 contest under two alternatives on the open-weight and smaller API models. \emph{Individual sampling}: each call instructs the model to act as one person matching the profile and answer survey questions (one option per question) at temperature 1.0; we draw $N{=}100$ personas per cell and form the per-(cell, question) distribution as the histogram of their choices. To cover all 280 cells per wave within budget, each persona answers a fixed random 10\% of the wave's questions (seeded per cell), so every cell--question pair is answered by $\approx$10 personas and coverage is complete at the cell level.

\emph{Log-probability readout}: each question is rendered with options labelled A, B, C, \dots{}; we take a single forward pass and apply a softmax over the first-token logits of the option letters (temperature 0, no sampling), the OpinionQA elicitation, collected for the open-weight models. Individual-sampling distributions rest on $\approx$10 personas per cell--question pair and should not be reused for absolute-accuracy claims; the contest is noise-immune at that $n$ (endpoints recomputed below) but TV errors are not. A second Mistral-7B individual-sampling run at doubled question coverage ($\approx$20 personas per cell--question, 14 waves, 244{,}933 cells) provides a dose--response check on the calibration itself: halving the target noise lowers the raw best-single rate from 77.8\% to 74.4\%---precisely the order-statistic deflation the endpoint machinery predicts---while the noise-matched index remains within the model family's band ($0.95$ at $n{\approx}10$ endpoints, $0.86$ at $n{\approx}20$). The raw rate tracks the noise dose; the calibrated conclusion does not move. Among the closed APIs, OpenAI's non-reasoning chat models do expose per-token log-probabilities (we verified that all option letters surface in the top-20 for a representative item), so the readout is feasible for GPT-4o-mini and GPT-4o; GPT-5.5 rejects the log-probability parameter, and the Anthropic API does not expose token probabilities. Both modes write the same JSONL schema as the aggregate pipeline, so the collapse contest and its calibration run unchanged. Two caveats on noise: individual-sampling and log-prob distributions are peakier than aggregate counts (one hard pick per persona; temperature-0 readout), inflating absolute TV distance but not the contest, which compares predictors of the \emph{same} bias vector and is immune to a shared per-cell noise floor. That immunity claim was verified rather than assumed: recomputing the calibration endpoints under the individual-sampling noise regime (each cell--question distribution a histogram of $\approx$10 hard picks, modelled as an $n{=}10$ multinomial draw) compresses the endpoints, exactly as coarser noise should (additive-truth best-single rises to 54.9\%, collapse-truth 79.1\%), yet the observed individual-sampling win rates ($\approx$78\%) still sit at the collapse end of the compressed scale (index $\approx$0.95). Within a persona's call, answers condition on its own earlier responses, as real survey respondents do. API individual-sampling runs cover the two-wave subset (W26, W34); open-weight runs cover all 15 waves (Extended Data Fig.~4 annotates each bar with the wave count entering its pooled contest).
\subsection*{Decomposition framework}
All quantities are defined per question $q$, and every vector operation takes place inside that question's response space: a question with $K$ answer options defines a $K$-dimensional simplex, and all distributions and difference vectors below live in $\mathbb{R}^K$ for that question; no comparison ever crosses questions. For a demographic cell $g$ we observe the human distribution $p_g$ and the model's simulated distribution $\hat p_g$, alongside the human population marginal $p_{\mathrm{pop}}$ and the model's own unconditioned (``Average American'') output $\hat p_{\mathrm{pop}}$, generated directly for every model. From these we form two kinds of difference vector. \emph{Steering} measures how conditioning on an identity moves a distribution away from the population: $s_g = p_g - p_{\mathrm{pop}}$ on the human side and $\hat s_g = \hat p_g - \hat p_{\mathrm{pop}}$ on the model side. \emph{Bias} measures how far the simulation sits from reality for the same cell: $e_g = \hat p_g - p_g$. Difference vectors sum to zero across options, so they live in the $(K{-}1)$-dimensional mean-zero subspace; magnitudes use total variation distance, $\tv(p,q)=\tfrac12\lVert p-q\rVert_1$, and directions use cosine similarity. The two are separated deliberately: composition is a claim about \emph{direction} (which way the distribution moves when identities combine), so the contest below is decided on cosines, whereas attenuation is a claim about \emph{magnitude} and is quantified separately by norm ratios. Three analyses use these objects in different spaces. The human additivity benchmark necessarily runs in steering space, since humans are the ground truth and have no bias vector; the headline model contest runs in bias space, because the question it answers is how the model's \emph{error} composes when identities are stacked; and the steering-space dominance analysis repeats the contest with $\hat s$ in place of $e$, requiring no ground truth at all, which confirms the result is not an artefact of referencing human data.
 
The composition contest is run independently for every pair cell $(A,B)$ and question. The model's own measured single-feature biases $e_A$ and $e_B$---outputs of the same model on the same wave and question---supply three candidate reconstructions of the realized pair bias $e_{AB}$:
\[
\underbrace{e_A + e_B}_{\text{additive}}
\qquad\text{versus}\qquad
\underbrace{e_A,\;\; e_B}_{\text{single-feature}},
\]
each scored by cosine similarity to $e_{AB}$. The best single is $\max_{d\in\{A,B\}}\Sim(e_{AB}, e_d)$; the cell is scored as \emph{collapsed} when the best single meets or beats the additive predictor, the pooled collapse share is the fraction of (cell, question) contests so scored, and the winning single defines the \emph{retained} (dominant) feature, $e_{\mathrm{dom}}$, used throughout the direction and magnitude analyses. Contests are decided by strict cosine comparison; exact ties do not occur at floating-point resolution (the code would score them for the single-feature predictor). At depth 3 the same construction competes the full additive sum $e_A+e_B+e_C$ against the three singles and, additionally, the three measured sub-pair biases, so a triple cell can be best explained by one feature, a pair, or the full combination (Extended Data Fig.~5).
 
The cosine contest separates hypotheses by direction. Symmetric additive composition aligns with $e_A+e_B$ even when attenuated---any $\lambda(e_A{+}e_B)$ with $\lambda<1$ points the same way, and its shrinkage is picked up by the magnitude ratios, not the contest---whereas exact single-feature retention \emph{or} any strongly lopsided weighting $\alpha e_A+\beta e_B$ with $\alpha\gg\beta$ aligns with a single-feature predictor. ``Collapse'' therefore names the family running from exact one-feature retention to heavily dominant weighting; the contest does not separate these sub-cases, and the magnitude analysis establishes that even the retained component is itself shrunk. Two norm ratios make this precise, and benchmark different things. The ratio $\|e_{AB}\|/\|e_{\mathrm{dom}}\|$ (median $0.95$) compares the realized pair bias to the retained single's own strength and requires no orthogonality assumption: exact retention ($e_{AB}=e_{\mathrm{dom}}$) would give exactly $1$, so a median below $1$ shows the kept feature is attenuated below its single-feature strength. The ratio $\|e_{AB}\|/\|e_A{+}e_B\|$ (median $0.59$) instead measures shrinkage relative to the additive prediction---a different and weaker benchmark, since the additive sum is the longer vector to begin with. Finally, raw cosines need a baseline: within a shared $(K{-}1)$-dimensional response space, even an \emph{irrelevant} bias vector has substantial expected alignment with $e_{AB}$, because model biases share question-level structure (the option geometry and the population-level component). The null calibration quantifies this floor by drawing, per cell, the bias of a random single profile from a dimension \emph{not} in the pair, on the same question; its cosine distribution (median 0.84) is the baseline against which the additive (0.94) and best-single (0.96) similarities are read (Extended Data Fig.~3).
 
The uncorrected additive predictor $e_A+e_B$ is exact only if the model has no population-level bias; the corrected predictor makes that assumption explicit and removes it. Write joint additivity in steering---$s_{AB}=s_A+s_B$ for humans, $\hat s_{AB}=\hat s_A+\hat s_B$ for the model---and expand the pair bias:
\[
e_{AB} \;=\; \hat p_{AB}-p_{AB}
\;=\; e_{\mathrm{pop}} + (\hat s_A - s_A) + (\hat s_B - s_B)
\;=\; e_A + e_B - e_{\mathrm{pop}},
\]
using $\hat s_d - s_d = e_d - e_{\mathrm{pop}}$, where $e_{\mathrm{pop}}=\hat p_{\mathrm{pop}}-p_{\mathrm{pop}}$ is the population-level model bias; the uncorrected predictor is the special case $e_{\mathrm{pop}}=0$. For triples the same expansion gives $e_A+e_B+e_C-2e_{\mathrm{pop}}$, and a sub-pair-based predictor carries its own single correction, $e_{XY}+e_Z-e_{\mathrm{pop}}$. The correction is asymmetric and touches only the additive arm: the collapse hypothesis is the bias-space statement $e_{AB}=e_{\mathrm{dom}}$, which contains no population term, so the single-feature predictor was already correctly specified and is unchanged---the corrected contest is strictly fairer to the additive hypothesis. (Collapse stated in output space, $\hat p_{AB}=\hat p_{\mathrm{dom}}$, differs from the bias-space statement only by the partner's human steering, $e_{AB}=e_{\mathrm{dom}}-s_{\mathrm{other}}$, a term small against pair biases---$\|s\|\approx0.04$--$0.09$ TV versus $0.24$--$0.35$---so the two readings coincide to first order.)
 
We re-ran the depth-2 and depth-3 contests with the corrected predictors as a robustness check, and the check is not vacuous, because $\lVert e_{\mathrm{pop}}\rVert$ is large---$0.8$--$1.0\times$ the pair-bias magnitude; the model cannot reproduce even the unconditioned marginal---so the correction nets out a population bias as large as the conditioning effect itself. This population bias is part of the single-feature miscalibration story: it is driven partly by the forced-choice refusal artefact (the flagship places 5.1\% in ``Refused'' versus 0.7\% in the real marginal) and partly by the same over-smoothing that compresses the single-feature hierarchy---the model's ``Average American'' is already a miscalibrated distribution before any identity is conditioned on. It also explains why the uncorrected bias-space contest is more generous to additivity than the steering-space one: $e_{\mathrm{pop}}$ is a component shared by every bias vector, and it is counted \emph{twice} in $e_A+e_B$ but only once in $e_{AB}$, lending the additive predictor spurious alignment; the correction removes exactly this double-counting. Using each model's own directly-generated $\hat p_{\mathrm{pop}}$, the corrected depth-2 collapse index spans \textbf{0.80--0.95}, statistically indistinguishable from the uncorrected shared-endpoint span of 0.80--0.93 (the self-calibrated indices of Table~\ref{tab:models} span 0.83--0.95), because the per-model win rates and the calibration endpoints ($40.3/84.3 \to 35.5/83.1$) shift together; raw win rates move a few points in both directions (e.g.\ GPT-4o-mini $78.4\to73.6$\%, Gemma-2-9B $75.3\to77.8$\%) and no conclusion changes. Table~\ref{tab:models} and the figures report the uncorrected contest as the primary analysis. The human-side contest of Fig.~\ref{fig:humancontest} needs no correction: it runs in steering space against the directly-measured survey population, so no model-population-bias term exists on the human side.
 
Three properties of the contest guard its validity. First, geometry: with $K=2$ options every difference vector lies on a single line, cosines are $\pm1$, and the contest is uninformative. Retaining the full response space including ``Refused'' lifts almost every item above this degenerate case---$K{=}2$ cells are 0.4\% of GPT-4o-mini's 335{,}979 pair cells, and excluding them leaves the collapse share at 78.4\%, unchanged. The share does decline with option count (83.9\% at $K{=}2$ to 52.4\% at $K{=}12$), because higher-dimensional response spaces give the additive predictor more room, but the pooled number is carried by the $K{=}3$--$5$ items that dominate the panel (92\% of cells). Second, margins: winning margins are small in absolute cosine terms (median $|c_{\mathrm{single}} - c_{\mathrm{add}}| = 0.023$; 34.7\% of contests within 0.01), which is precisely why raw cosines and raw win rates are never interpreted directly: the calibration endpoints are computed under the identical contest, margins and all (see \emph{Statistical inference}). Third, the maximum: the collapse statistic rests on the \emph{max} over the two singles, which enjoys a mechanical post-hoc advantage under noise; both singles simultaneously beat the additive predictor in only 0.5\% of cells, so the statistic tracks the best single rather than any uniform superiority of singles, and the max's inflation is exactly what the endpoint calibration absorbs. Fourth, magnitude: win rates are flat across deciles of $\lVert e_{AB}\rVert$ (77.3--81.0\%), excluding the smallest-norm cells ($\lVert e_{AB}\rVert<0.05$; 1.5\% of cells) moves the pooled share by 0.1 point, and the zero-norm cosine guard never fires on real cells (0 of 363{,}714), so low-signal cells do not carry the result. The random-unrelated null of Extended Data Fig.~3a is constructed per pair cell--question by drawing one single-feature bias vector uniformly at random (fixed seed) from the same question's measured singles whose dimension differs from both of the pair's dimensions; subtracting the shared population-bias component $e_{\mathrm{pop}}$ from both sides collapses this null from median cosine 0.84 to 0.28, confirming that the floor is essentially the shared population bias.

\subsection*{Coefficient decomposition and identifiability}
For Fig.~\ref{fig:humancontest} and Extended Data Fig.~2 we solve, per pair cell and
question, the two-parameter least-squares problem $t \approx \alpha u +
\beta v$, where $(t,u,v)$ are the realized pair vector and its two
single-feature parts: steering vectors $(s_{AB}, s_A, s_B)$ on the human
side (cells with $n\ge100$) and bias vectors $(e_{AB}, e_A, e_B)$ on the
model side---the two panels of Fig.~\ref{fig:humancontest} live in these two different
spaces, as labelled, because humans have no bias vector to decompose.
Cells whose basis vectors are near-collinear ($|\cos(u,v)|>0.95$), where
the coefficients are unidentifiable, are excluded ($\sim$6\% of cells).
Collinearity short of that gate still matters: model bias vectors share
the population-level component $e_{\mathrm{pop}}$, so the two regressors
are often correlated, and correlated regressors can by themselves
produce lopsided $(\alpha,\beta)$ estimates. The ridge is not that
artifact. Stratifying the flagship's cells by basis angle, the median
budget is $\alpha{+}\beta = 1.02$ in \emph{every} stratum, and in the
best-conditioned stratum ($|\cos(e_A,e_B)|<0.3$; 26{,}282 cells), where
least squares is well identified, the median dominant share remains
$0.82$ with 58\% of cells above $0.75$. Conditioning inflates the
extremity of the split (dominant share $0.94$ in the most collinear
stratum, $0.7\le|\cos|<0.95$) but leaves the budget untouched---and the
budget is the collapse claim. The human coefficient cloud is broadened
by ground-truth sampling noise, which is unbiased for the centre; the
reported medians are the estimand. The model panel has no such excuse:
its ridge lies along $\alpha{+}\beta{=}1$, not around $(1,1)$.

\subsection*{Noise floor and noise corrections}
Every accuracy statistic compares a model distribution to a human distribution that is itself an estimate: a cell of $n$ respondents yields an empirical $\hat p_n$, not the true $p$, so even a perfect simulator scores nonzero TV against it---irreducible error at the level of whole distributions. Uncorrected, this makes small cells look artificially hard and, worse, confounds every cross-depth comparison, since deeper intersections have systematically smaller $n$. All accuracy statistics are therefore referenced to a per-cell noise floor---the TV a perfect simulator would be expected to score against that cell, given its $n$ and the question's response distribution---rather than to zero. The floor is instrumented twice. Empirically: for every human cell with $n\ge40$ (each half then clears the $n\ge20$ gate), we split the respondents into two random halves (twice, seeded) and compute the TV between the halves' survey-weighted distributions, per question---both halves estimate the same $p$, so their TV is pure sampling noise, and the construction automatically absorbs weighting and panel-design effects. Analytically: a closed-form per-cell expectation, available for every cell and question and cheap to subtract everywhere, which the empirical floor certifies.

The closed form combines three ingredients: TV is a sum of coordinatewise absolute deviations, $\tv(\hat p_n,p)=\tfrac12\sum_i|\hat p_{n,i}-p_i|$; each empirical frequency is a multinomial proportion, so $\hat p_{n,i}-p_i \approx \mathcal N\!\big(0,\,p_i(1-p_i)/n\big)$; and $E|Z|=\sigma\sqrt{2/\pi}$ for a centred normal. Linearity of expectation---no independence across options required---then gives
\[
E\big[\tv(\hat p_n, p)\big] \;\approx\; \tfrac12\sqrt{2/(\pi n)}\,\sum_i \sqrt{p_i(1-p_i)}.
\]
The split-half expectation is twice this at the cell's full $n$: one $\sqrt2$ because each half has half the respondents, one because two noisy estimates are compared rather than one against the truth. Because this expectation is subtracted inside every excess-error statistic, any error in it propagates into every accuracy number we report; we therefore certify it across its full range of use with two complementary checks. Observed split-half noise on real cells is 1.01--1.02$\times$ this prediction, certifying the formula against real-panel effects (weighting, design)---but that check requires $n\ge40$, so it cannot reach the smallest cells, where the normal approximation is least reliable. Multinomial simulation---drawing from known distributions, so the exact expectation is computable---covers precisely that regime, bounding the closed form within 0--4\% of it down to $n=20$, erring conservative: where the closed form is off, it overstates the floor, so excess error is understated. For scale, a perfect simulator is expected to score $\tv\approx0.15$ against an $n{=}20$ cell on a uniform four-option item, and $\approx0.07$ at $n{=}100$. Two further design checks: the recombination analysis of Fig.~\ref{fig:decomp} replicates cross-fitted---singles and population estimated on one random respondent half, the pair target on the other---at mean TV 0.106, below the half-sample split-half noise of 0.134, so the low recombination error is not shared-sampling-noise leakage; and the depth growth of human distinctiveness is robust both to survey-weight design effects ($2.87\times$ under raw $n$ versus $2.86\times$ under Kish $n_{\mathrm{eff}}=(\sum w)^2/\sum w^2$; mean design effect 1.8--2.0) and to cell composition (recomputed within each of the 35 dimension-triple families, growth persists in every family: median $2.9\times$, range $2.1$--$4.6\times$).

\emph{Excess error} subtracts this per-cell expectation from the observed TV: the model's error above what a perfect simulator would score on that cell. Because absolute values do not decompose ($E|a+\mathrm{noise}|\ne|a|+E|\mathrm{noise}|$), the subtraction is first-order and, where true error dominates the noise, over-corrects---excess errors are, if anything, understated. The distinctiveness analysis (Extended Data Fig.~1) demands exactness instead, since its claim compares depths whose cells shrink systematically, and any residual noise bias would masquerade as growing distinctiveness. It therefore uses squared Euclidean distance, whose expectation decomposes exactly---$E\lVert \hat p_g - \hat p_{\mathrm{pop}}\rVert^2 = \lVert p_g - p_{\mathrm{pop}}\rVert^2$ plus the two coordinatewise variance sums---so subtracting their plug-in estimates,
$\lVert \hat p_g - \hat p_{\mathrm{pop}}\rVert^2 - \sum_i \hat p_{g,i}(1-\hat p_{g,i})/n_g - \sum_i \hat p_{\mathrm{pop},i}(1-\hat p_{\mathrm{pop},i})/n_{\mathrm{pop}}$,
yields an estimator unbiased for the true squared distinctiveness at every depth. (The cell sits inside the population sample; the induced covariance over-subtracts by a small, depth-independent amount of order $\sum_i p_{g,i}(1-p_{g,i})/n_{\mathrm{pop}}$, understating levels without touching the growth.)

\subsection*{Human additivity contest and contest calibration}
The model contest presupposes that the additive vector is the \emph{right} target---that real intersectional subgroups compose additively. This subsection tests that premise rather than assuming it, and calibrates the contest itself. The human-side contest mirrors the model contest but runs in steering space, the only space in which humans can be interrogated (they are the ground truth and have no bias vector): for every pair cell with $n\ge100$ respondents on a question---population, both singles, and the pair all observed---the additive predictor $s_A+s_B$ competes against the better single steering vector for cosine similarity to the cell's real steering $s_{AB}$, and the model is scored on the identical cells in its own steering space, so the human--model comparison is not confounded by cell composition. Additivity is defined in probability space by deliberate choice: the vector machinery (sums, cosines) requires a linear rule, and probability-space additivity is the first-order approximation to the canonical no-interaction (log-linear/independence) model, coinciding with it to first order in the small-tilt regime we observe (per-question steering TV $\approx0.04$--$0.09$); because the human target is defined empirically and the models fail to match it under either rule, the choice of composition rule does not affect the conclusion. Even at $n\ge100$, ground-truth sampling noise biases the contest toward the single-feature predictor---the best single is a maximum and harvests noise, while the additive predictor carries the summed noise of two estimates---so the raw human win rate cannot be read against an ideal 100\%. We therefore measure the attainable ceiling directly: synthetic pair cells whose true distribution is exactly $p_{\mathrm{pop}}+s_A+s_B$ (elementwise $\max(\cdot,0)$, then renormalised; the clip activates in 15.1\% of pair cells at $n\ge100$), observed through one multinomial draw at the cell's actual $n$---a perfectly additive population seen through exactly the noise of the real one---and scored by the same contest. Human additivity is judged against this ceiling. The model-side ceiling is constructed identically in the model's steering space: synthetic cells $\hat s^*_{AB}=\hat s_A+\hat s_B+\eta$, with $\eta$ drawn from the model's measured run-to-run deltas (option-count matched), scored by the same contest on the same cells (29.7\%).

The headline bias-space statistic is calibrated by the same measure-rather-than-assume logic: the contest's built-in biases (the maximum over singles, the small margins, the option-count geometry, the correlations among a model's own single biases) resist closed-form correction, so the endpoints are measured by replaying the identical contest on synthetic truths built under each pure hypothesis. Synthetic pair biases are $e^*_{AB}=e_A+e_B+\eta$ (additive truth) and $e^*_{AB}=e_{\mathrm{dom}}+\eta$ (collapse truth; $e_{\mathrm{dom}}$ is the larger-norm of the cell's two measured single-feature biases), with $\eta$ drawn from measured run-to-run deltas (all pairs of the four repeated runs (six pairs), scaled by $1/\sqrt{2}$ because a delta between two runs carries both runs' noise, and matched on option count because noise geometry depends on $K$); run-to-run stochasticity is the model-side analogue of sampling noise, measured rather than posited. The collapse index then linearly rescales an observed win rate so that the additive-truth endpoint is 0 and the collapse-truth endpoint is 1, converting raw rates---incomparable across models, paradigms, and depths---into positions on a common per-model scale (see \emph{Statistical inference}). The construction extends to depth 3 (synthetic triples from three measured single biases; the sub-pair predictors carry their own noise draw, because real sub-pair predictors are themselves measured model outputs, and noiseless rivals would flatter them). The depth-3 endpoints show why raw shares mislead: a perfectly additive model attains an additive share of only 67.2\%---its predictor sums three noisy vectors and faces six noise-harvesting rivals---while a pure collapser still reaches 6.6\%, so the observed 7.8--10.0\% sits at the pure-collapse floor; per-model endpoints from each model's own noise give Index$_{\mathrm{d3}}$ 0.92--0.98 (Table~\ref{tab:models}).

The index itself carries little estimation uncertainty: propagating both sources---a wave-cluster bootstrap on the observed share and a jackknife over the six repeat-pair noise sets on the endpoints---gives the flagship's index a 95\% CI of $\pm0.02$, with endpoint standard errors of 0.15 and 0.06 points; endpoint estimation error is negligible against the size of the collapse effect. Endpoint representativeness beyond the repeat wave rests on the option-count matching of $\eta$; the dose--response check above and the option-count-stratified indices in Results bound that assumption empirically. Two checks establish that the index does not hinge on the noise model. The per-model self-calibration (Index$_{\mathrm{self}}$, $0.83$--$0.95$), which substitutes each model's own \emph{measured} $\eta$ rather than a scaled surrogate, is the primary check: it removes the noise model entirely. The scaling sweep then bounds the surrogate's influence: rescaling the noise magnitude by $0.5\times$ leaves GPT-4o-mini's index at $0.86$ (unchanged from $1\times$), and even doubling it---far beyond the measured run-to-run spread---only softens it to $0.74$, still firmly on the collapse side.

\subsection*{Collapse direction, permutation null, steering dominance}
For each collapsed pair cell (best single $\ge$ additive), the winner is the dimension with the larger bias cosine; the human-dominant dimension is the one whose real single subgroup is farther (TV) from the population on that question. Because the pair cell's respondents also sit inside its parent single cells, predictor and target share sampling noise (leakage). This sharing does not drive the human result: stratifying the human contest by $\max(n_{AB}/n_A, n_{AB}/n_B)$ \emph{within} cell-size bands---which separates leakage from the target-noise gradient that dominates the raw stratification---bounds the residual association at $0$--$5$ points (38.6/37.9/40.6\% across leakage tertiles at $n_{AB}\in[100,200)$), and in the lowest-leakage, largest-cell stratum, where the single-feature estimates are most independent of the target, the additive win rate is $48.7\%$---at the perfectly additive composer's ceiling. 

Collapse rates themselves are tightly banded across all 21 dimension pairs (75.5--83.1\%), so retention deltas are not driven by which pairs enter the collapsed set; at the value level the deltas are carried by minority values (Black $-19.0$, Hispanic $-22.3$, Asian $-19.0$, mixed-race $-19.3$, atheist $-20.7$, Jewish $-23.0$ points, against White $+16.3$, Protestant $+6.4$, Roman Catholic $+3.9$). Matched-wave generation contrast (W26$+$W34, identical cells): prior-generation pooled deltas race $-7.3$, religion $-3.0$, gender $+2.3$; GPT-5.5 $-6.2$, $-1.7$, $-1.0$; Claude Sonnet 5 race $-8.0$, religion $-3.1$; the sharper pooled all-wave contrasts (race $-9.5$; gender $+9$ to $+17$) partly reflect wave composition rather than the models. The pairing-level rates of Fig.~\ref{fig:direction}a aggregate, for each (dimension, partner) pairing with $\ge$2{,}000 contested cells, the share of contests in which the dimension is human-dominant (abscissa) and the share in which the model retains it (ordinate). The permutation null shuffles human-dominant labels within (wave $\times$ dimension-pair) strata (500 permutations), preserving both marginal rates. Steering-space dominance repeats the contest with steering vectors $\hat s$ in place of bias vectors, requiring no ground truth. 

\subsection*{Base-versus-tuned contrast}
To locate the sensitive-dimension suppression in the training pipeline we ran three conditions on Llama-3.1-8B, over the 15 waves with full base-condition coverage, with Mistral-7B replicating the A-versus-B weight contrast on the two waves where its base-prompt instruct condition is available: (A)~the base pretrained checkpoint under the base completion prompt; (B)~the instruction-tuned checkpoint under the same base prompt; (C)~the instruction-tuned checkpoint under its chat template. A vs.\ B isolates the effect of tuning the \emph{weights} at fixed prompt; B vs.\ C isolates the \emph{prompt format} at fixed weights. Throughout, ``preference-tuning'' denotes the full post-training pipeline bundled into the public instruct checkpoint---supervised fine-tuning and preference optimisation (RLHF/DPO) together---which we cannot separate; the contrast distinguishes the entire post-training stage from pretraining, not RLHF from SFT. For each dimension we report $\Delta = (\text{keep rate}) - (\text{human-dominance rate})$ in percentage points, where the keep rate is the share of collapsed pairs that retain that dimension and the human-dominance rate is the share in which the dimension is the human-dominant one; a negative $\Delta$ is suppression. The base condition recasts the task as a document-completion: a header describing a database of survey records, two few-shot exemplar records on demographically neutral items (weather-forecast frequency and commuting mode) that teach the output format without touching any identity dimension, then the target record ending at an open bracket (verbatim prompt in the released code). Output validity is comparable across conditions: base-checkpoint completions parse at 100.0\% (Llama base prompt) and 99.9--100.0\% (instruct weights under base prompt; Mistral base 100.0\%), and the one lower-validity condition (Mistral instruct under the base prompt, 78.3\%) loses cells uniformly across dimensions (72--82\%, no dimension-selective gap), so condition contrasts compare like-with-like cells rather than a parse-selection artefact. Differences and their $95\%$ CIs use the wave-cluster bootstrap ($B=10{,}000$). To guard against reading a null as evidence of no effect, we report each contrast's minimum detectable effect: with $\mathrm{SE}$ the bootstrap standard error of the difference, the two-sided shift detectable at $80\%$ power and $\alpha=0.05$ is $(z_{0.975}+z_{0.80})\,\mathrm{SE}=2.80\,\mathrm{SE}$. For race this is $5.9$ points (observed B$-$A $=+0.6$) and for religion $2.7$ points (observed $-0.6$), both well below the suppression magnitudes themselves ($-10.8$, $-5.3$). The base checkpoint demonstrably conditions on demographics rather than ignoring them: on shared cells its per-dimension bias magnitudes match the instruct checkpoint under the identical prompt (mean TV 0.24--0.27 for both), and its differentiation between a dimension's value outputs (mean pairwise TV 0.19--0.23) exceeds the human reference (0.07--0.15).

\subsection*{Entropy analysis}
For matched cells, $\Delta H = H(\hat p_g) - H_{\mathrm{MM}}(p_g)$, with
the Miller--Madow correction $H_{\mathrm{MM}} = H_{\mathrm{plugin}} +
(K-1)/(2n)$ applied to the human side. The refusal artifact is assessed
by removing refusal options and renormalising both distributions.
Robustness: $\Delta H$ is positive with wave-cluster CIs excluding zero at
every depth ($+0.18/+0.21/+0.23$ nats at depths 1/2/3), under restriction
to cells with $n\ge100$, normalised by $\log K$, under a Herfindahl
impurity measure, and after refusal exclusion ($+0.09/+0.11/+0.13$ by
depth), and is monotone in depth.

\subsection*{Statistical inference}
This study was not preregistered; all analyses are exploratory, and the
paper's claims are estimation-based rather than significance-based:
headline quantities are shares and mean distances reported with
confidence intervals and read against \emph{measured} reference
points---the calibration endpoints and the human sampling-noise
floor---rather than against zero. $P$ values appear in exactly two
analyses (the permutation test for retention--relevance agreement and
the Spearman correlations of the typicality analysis); both are reported
exactly, with sidedness stated.

\textbf{Unit of analysis and clustering.} Every pooled statistic is a
ratio of sums over (profile $\times$ question) cells. Cells are not
independent: cells within a wave share the question set, the fielding
period, and the respondent sample. The wave is therefore the primary
cluster, and all headline intervals are 95\% percentile intervals from a
nonparametric cluster bootstrap that resamples whole waves with
replacement ($B=10^4$), recomputing the pooled ratio on each resample.
Direction statistics, whose estimands vary by dimension pair, cluster on
(wave $\times$ dimension-pair) strata instead ($B=2{,}000$). For the
two-wave frontier models, wave resampling is degenerate; their intervals
cluster on questions---the dominant residual dependence within a
wave---using the 143 question clusters of W26$+$W34, and are marked
$^{\ddagger}$ in Table~\ref{tab:models}.

\textbf{Small-cluster robustness.} Fifteen waves is a modest number of
clusters, so the wave bootstrap is checked against anticonservatism two
ways. A wild-cluster bootstrap (Rademacher signs on wave-level
deviations, $B=10^4$) brackets every headline within its reported
interval (GPT-4o-mini depth-2 collapse $[77.8, 79.0]\%$; depth-3
additive $[7.4, 8.2]\%$; the same holds on Gemma-2, Mistral, and Llama).
Leave-one-wave-out re-estimation moves no headline point by more than
0.2 points (GPT-4o-mini collapse $[78.2, 78.6]\%$): no single wave
drives any result.

\textbf{Permutation null.} Agreement between the retained and the
human-dominant dimension has no intrinsic 50\% chance level, because
dominance rates differ across dimensions. The null distribution is built
by shuffling human-dominant labels within (wave $\times$ dimension-pair)
strata (500 permutations), which preserves both marginal rates; the
observed agreement exceeds every permuted value for every model, so the
one-sided permutation $P$ is reported at the resolution of the test,
$P\le1/501$.

\textbf{Minimum detectable effects.} Null contrasts (base-versus-tuned)
are reported with their minimum detectable effect rather than read as
evidence of absence: with $\mathrm{SE}$ the wave-cluster bootstrap
standard error of the contrast, the two-sided shift detectable at 80\%
power and $\alpha=0.05$ is $(z_{0.975}+z_{0.80})\,\mathrm{SE}=2.80\,
\mathrm{SE}$.

\textbf{Split-sample confirmation.} Every headline number is recomputed
separately on odd and even waves (by chronological order); all replicate
(e.g.\ depth-2 collapse 77.8\%/78.9\%; depth-3 additive 7.9\%/7.7\%;
race retention delta negative in every half-sample, $-4.3$ to $-8.7$
points). One caveat: the ATP is a panel, so the two halves are disjoint
in waves and questions but not in people---84\% of the smaller half's
respondents also appear in the other half (mean pairwise wave overlap
60\%). The split therefore demonstrates stability across question sets,
topics, and fielding periods, not respondent-level independence; because
every wave contributes entirely different questions, the compared cells
are nonetheless distinct measurements, and all cluster intervals should
be read as clustering over question-set and fielding draws rather than
over independent samples of people.

\textbf{Statistics and reproducibility.} No statistical method was used
to predetermine sample size: ground-truth coverage is set by the survey
instrument (every demographic cell with $n\ge20$ valid respondents), and
model coverage by compute budget, stated per model in
Table~\ref{tab:models}. No data were excluded beyond the stated gates.
Randomisation and blinding are not applicable: the study is a secondary
analysis of existing survey microdata combined with deterministic
scoring of model outputs, with no experimental group assignment. No
distributional assumptions underlie the reported intervals: all
inference is by resampling (cluster bootstrap, wild-cluster bootstrap,
permutation) or exact multinomial theory (the noise floors), and every
resampling procedure is seeded. Analyses used Python (NumPy/SciPy); the
full scoring pipeline, seeds, and per-cell outputs will be released
(see Code availability).

\subsection*{Wave sets and exclusions}
The panel comprises 15 waves. (i) All pooled statistics---spine, collapse shares, collapse direction, bootstrap and permutation inference, the noise floor, and every number in the abstract and Table~\ref{tab:models}---use the \textbf{15 waves} common to every full-coverage model; the two-wave models of (v) enter no cross-model pooling); Fig.~\ref{fig:direction}a pools the five full-coverage models on the same 15 waves. (ii) The per-topic analysis (Fig.~\ref{fig:direction}b) shows all \textbf{15 topics}; (iii) The reliability analysis uses \textbf{one wave} (W26) run four times end-to-end. (iv) the elicitation comparison (Extended Data Fig.~4) pools each paradigm's available waves as labelled, a deliberate exception to this convention. (v) GPT-5.5 and Claude Sonnet 5 cover a \textbf{two-wave} subset (W26, W34); their pooled statistics are computed on those waves and they are flagged in every display where they appear. (vi) The base-versus-tuned contrast uses the \textbf{15 waves} with full base-condition coverage.

\subsection*{Use of large language models}
Large language models were used in this work in two distinct roles, and
we disclose both. First, as the objects of study: the eight evaluated
models generated the simulated distributions analysed throughout.
Second, as research assistance: Claude (Anthropic) was used, under
continuous author direction, to assist with analysis and pipeline code, figure preparation, and
language editing of the manuscript. All analyses were specified by the
authors, every reported statistic was verified against the released code
and data, and the authors take full responsibility for the content. We
note for transparency that the assisting model belongs to the same model
family as two of the evaluated systems (Claude Haiku 4.5 and Claude
Sonnet 5); the evaluation pipeline, scoring rules, and conclusions were
fixed by the authors and apply identically to all eight models. No
language model is an author.

\subsection*{Ethics}
No new human data were collected. All human data are secondary analyses of de-identified, publicly available Pew ATP microdata, originally collected by Pew Research Center under its panel consent protocols.

\subsection*{Data availability}
All 21.1M simulated response records (15.7M primary pipeline; 2.1M
base-versus-tuned conditions; 1.3M log-probability readout; 1.8M individual
sampling; 0.2M reliability, prompt-sensitivity, and position controls),
together with all derived analysis artifacts and aggregated human response
distributions, are available on Zenodo at
\url{https://doi.org/10.5281/zenodo.21267988}. Respondent-level ATP microdata
cannot be redistributed; the waves are available from Pew Research Center
(free registration), and we accessed processed versions via the OpinionQA
release\citep{santurkar2023whose}. The deposit documents how to place the
microdata for full recomputation.

\subsection*{Code availability}
All generation, scoring, analysis, and figure code is included in the Zenodo
deposit and maintained at
\url{https://github.com/VRennard/Intersectionality-and-Synthetic-Identities}.
Every figure and statistic in the paper can be regenerated from the deposit
with \texttt{run\_all\_figures.sh} (Python~3 with numpy, pandas, matplotlib).

\section*{Author contributions}
Conceptualization: VR;
Methodology : VR;
Software : VR, CX;
Validation : VR;
Formal Analysis : VR;
Investigation : VR;
Data Curation : CX, VR;
Writing - Original : VR;
Writing - Review : CX;
Visualization : VR, CX;
Supervision : VR;
Project Administration : VR;
Funding Acquisition : VR

\section*{Competing interests}
The authors declare no competing interests.

\section*{Additional information}
\textbf{Correspondence and requests for materials} should be addressed to Virgile Rennard (ORCID: 0009-0001-8951-0482)

\bibliographystyle{unsrtnat}
\bibliography{refs}

\section*{Extended Data}

\begin{figure}[p]
\centering
\includegraphics[width=0.72\textwidth]{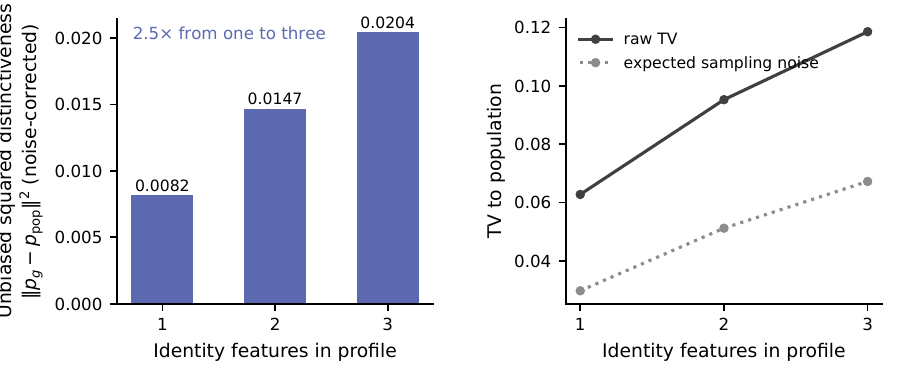}
\caption*{\textbf{Extended Data Fig.~1 $|$ Real subgroups grow more
distinctive as identities intersect.} Noise-corrected squared
distinctiveness of human cells by profile depth (0.0082/0.0147/0.0204),
with raw TV and expected noise for reference.}
\end{figure}

\begin{figure}[p]
\centering
\includegraphics[width=\textwidth]{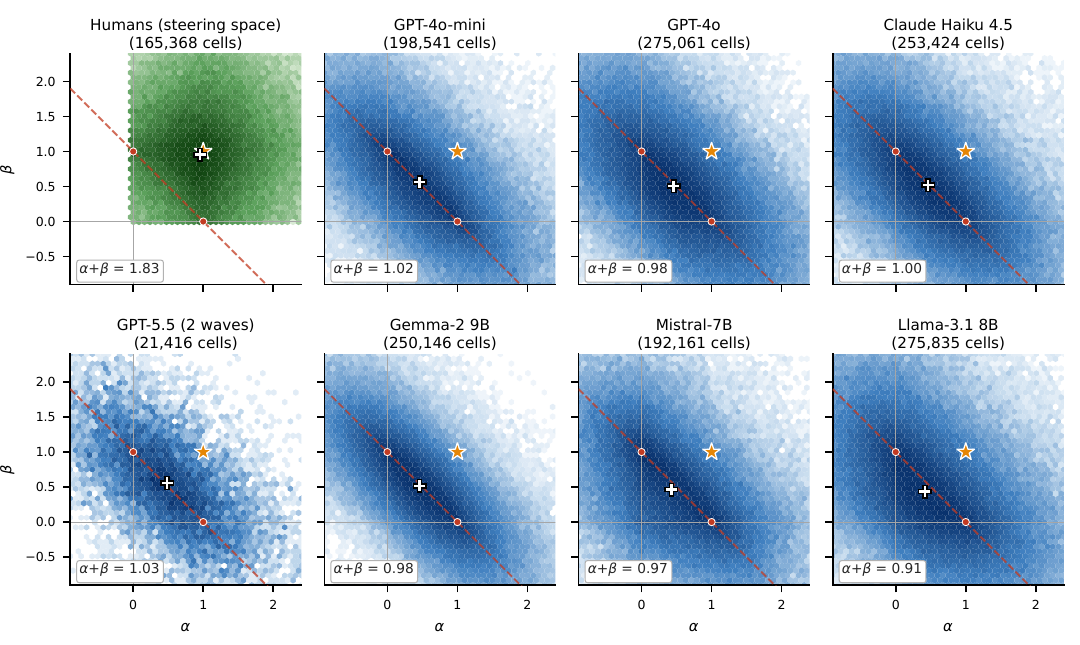}
\caption*{\textbf{Extended Data Fig.~2 $|$ Per-model replications.} The
coefficient decomposition of Fig.~\ref{fig:humancontest}, per model: each panel decomposes the
model's realized pair-bias vectors as $\alpha$ times the $A$-part plus
$\beta$ times the $B$-part (log-scale hexbins; white cross, median; star,
perfect composition; dashed, $\alpha{+}\beta{=}1$). Humans (green, steering
space) centre on composition with median budget 1.83; all seven models ride
the $\alpha{+}\beta{=}1$ ridge with median budgets 0.91--1.03---one
identity's worth of weight, in every architecture and both model
generations (GPT-5.5 on its two-wave subset). Per-model collapse shares and
confidence intervals are in Table~1 and Extended Data Fig.~5.}
\end{figure}

\begin{figure}[p]
\centering
\includegraphics[width=0.9\textwidth]{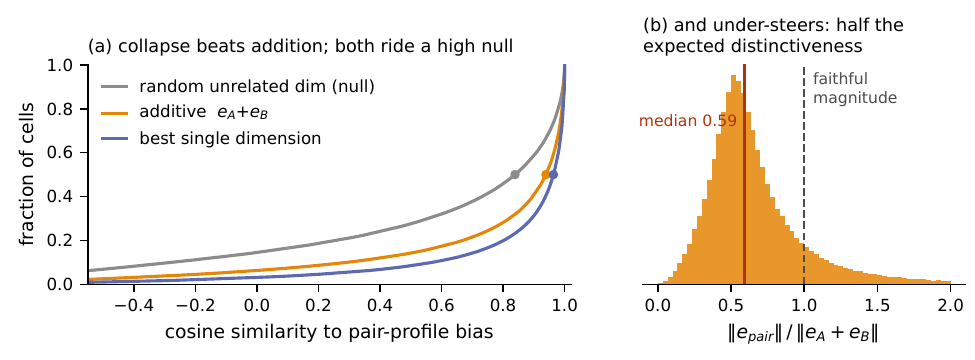}
\caption*{\textbf{Extended Data Fig.~3 $|$ Null-calibrated similarity and
magnitude shrinkage.} Cumulative cosine distributions of the
random-dimension null (median 0.84), additive prediction (0.94) and best
single (0.96), and the distribution of
$\lVert e_{\mathrm{pair}}\rVert/\lVert e_A{+}e_B\rVert$ (median 0.59),
for the flagship model.}
\end{figure}

\begin{figure}[p]
\centering
\includegraphics[width=0.85\textwidth]{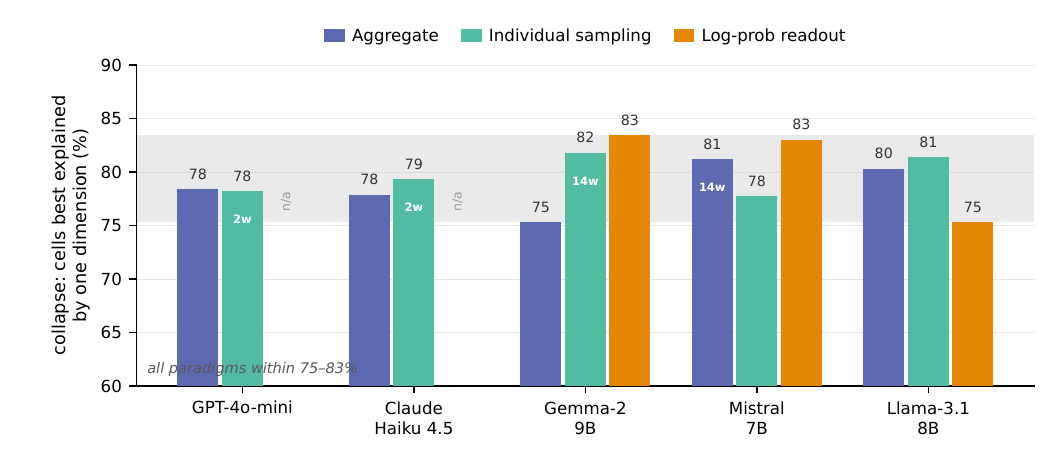}
\caption*{\textbf{Extended Data Fig.~4 $|$ Elicitation invariance.}
Depth-2 best-single win rate for five model families under aggregate,
individual sampling, and log-probability elicitation; every (model
$\times$ paradigm) cell lies within 75.3--83.4\%.}
\end{figure}

\begin{figure}[p]
\centering
\includegraphics[width=0.6\textwidth]{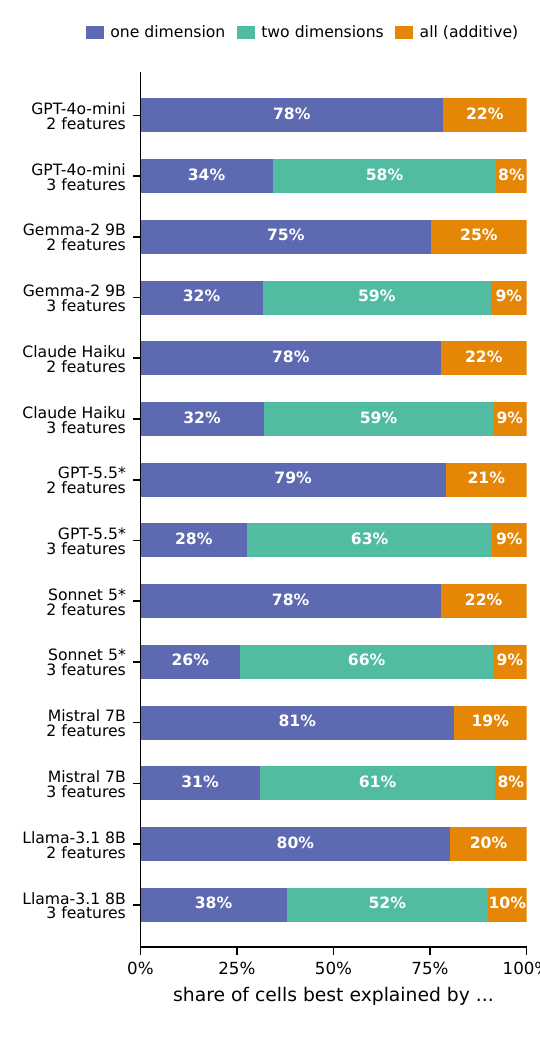}
\caption*{\textbf{Extended Data Fig.~5 $|$ Predictor shares by depth and
model.} Share of cells best explained by one dimension, two dimensions,
or the full additive combination, for the seven models with
three-feature data; the additive share falls from 19--25\% (two
features) to 7--10\% (three).}
\end{figure}

\begin{figure}[p]
\centering
\includegraphics[width=0.8\textwidth]{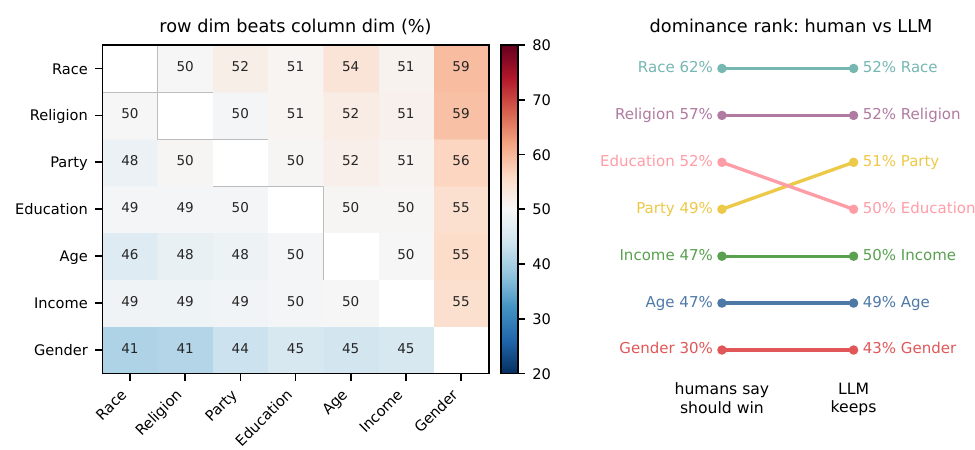}
\caption*{\textbf{Extended Data Fig.~6 $|$ Pairwise dominance matrix.}
Share of collapsed cells in which the row dimension's bias explains the
pair, for all dimension pairs, pooled over the five full-coverage
models, with each dimension's rank by human dominance versus model keep
rate.}
\end{figure}

\begin{figure}[p]
\centering
\includegraphics[width=0.75\textwidth]{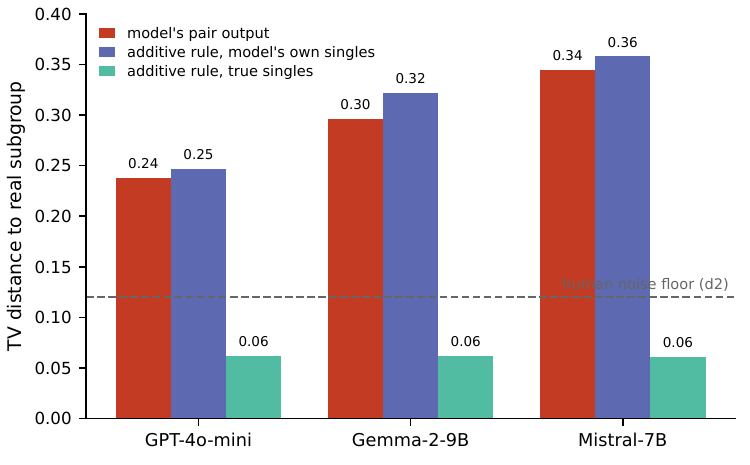}
\caption*{\textbf{Extended Data Fig.~7 $|$ Accuracy decomposition,
per-model means.} Native pair error versus the additive rule on the
model's own singles versus the same rule on true singles (GPT-4o-mini
0.238/0.247/0.062; Gemma-2-9B 0.296/0.322/0.062; Mistral-7B
0.345/0.358/0.061), against the depth-2 noise floor.}
\end{figure}

\begin{table}[p]
\centering
\begin{tabular}{lccc}
\toprule
 & Depth 1 & Depth 2 & Depth 3\\
\midrule
Model run-to-run TV (four repeats)        & 0.119 & 0.120 & 0.120\\
Human split-half TV (weighted, same wave) & 0.096 & 0.134 & 0.159\\
\bottomrule
\end{tabular}
\caption*{\textbf{Extended Data Table~1 $|$ Reliability.} Flagship
run-to-run total variation across four repeated end-to-end runs of W26,
by profile depth, against the weighted human split-half noise on the
same wave. The across-run spread of the aggregate per-depth error is
$\le$0.001 TV, so prompt stochasticity cannot account for any reported
effect.}
\end{table}

\begin{table}[p]
\centering
\begin{tabular}{lcc}
\toprule
 & Weighted & Unweighted\\
\midrule
Depth-2 collapse (best-single share) & 78.4\% & 78.3\%\\
Depth-3 additive share               & 7.8\%  & 8.3\%\\
Direction agreement (flagship)       & 57.9\% & 58.4\%\\
\midrule
Ground-truth mean TV, weighted vs unweighted: & &\\
\quad population / single / pair cells & \multicolumn{2}{c}{0.029 / 0.040 / 0.053}\\
\bottomrule
\end{tabular}
\caption*{\textbf{Extended Data Table~2 $|$ Survey-weight sensitivity.}
Headline numbers recomputed with survey-weighted (primary) versus
unweighted ground-truth distributions, and the mean TV between the two
ground-truth bases by cell type---4--6$\times$ smaller than any model's
error.}
\end{table}

\begin{table}[p]
\centering
\begin{tabular}{lccc}
\toprule
$\Delta H$ variant & Depth 1 & Depth 2 & Depth 3\\
\midrule
Raw (Miller--Madow, nats)      & $+0.180$ [0.159, 0.201] & $+0.206$ [0.186, 0.225] & $+0.234$ [0.216, 0.251]\\
Cells with $n\ge100$           & $+0.180$ [0.159, 0.201] & $+0.200$ [0.180, 0.220] & $+0.222$ [0.202, 0.243]\\
Normalised, $\Delta H/\log K$  & $+0.127$ [0.112, 0.143] & $+0.146$ [0.132, 0.160] & $+0.168$ [0.156, 0.180]\\
Herfindahl impurity difference & $+0.075$ [0.062, 0.087] & $+0.092$ [0.081, 0.104] & $+0.111$ [0.102, 0.121]\\
Refusal options excluded       & $+0.088$ [0.069, 0.105] & $+0.107$ [0.088, 0.124] & $+0.126$ [0.108, 0.142]\\
\bottomrule
\end{tabular}
\caption*{\textbf{Extended Data Table~3 $|$ Within-group dispersion
robustness.} Simulated-minus-real dispersion for the flagship by profile
depth, under five estimators: $\Delta H$ is simulated entropy minus
Miller--Madow-corrected real entropy (positive = simulation more
dispersed); rows restrict to large cells, normalise by option count,
substitute a Herfindahl impurity difference, and exclude refusal options.
Brackets, 95\% wave-cluster bootstrap CIs (15 waves). The effect is
positive under every estimator and monotone in depth.}
\end{table}

\end{document}